\documentclass{article}
\usepackage{amssymb}
\usepackage{amsfonts}
\usepackage{amsmath}
\usepackage{graphicx}

\input{tcilatex}
\begin{document}

\title{Is Brain in a Superfluid State? Physics of Consciousness.}
\author{Benoy Chakraverty **}
\date{}
\maketitle
\tableofcontents

\begin{abstract}
We set up the human brain as a quantum field of Information in the cognitive
functional space of the mind. To this end, a quantum operator $s$ is
introduced which will create information like particle (called \textit{infons%
}) and generate a coherent macroscopic\ information field. \textit{This
operator \ represents self and reflects our genetic identity.}The non-zero
average of this non-hermitian operator,denoted by $\left\langle
s\right\rangle $ is defined \ as the cognitive self usually referred to as
the first person $\mathcal{I}$ \ in our everyday life. A local field
operator $\psi _{i}$ is defined that generates infons at neuronal synaptic
sites $i.$ We impose the identity of synaptic self $\left\langle \psi
\right\rangle $ with the cognitive self $\mathcal{I}$. We establish
consciouness as the causal cognitive response function of brain or a
susceptibility to the external world. We show that at the emergence of $%
\left\langle s\right\rangle $, self-consciousness rises out of
consciousness.This is reflected precisely by divergence of the
susceptibility function ; an infinitesimal perturbation due to external
world becomes an incredibly intense cognitive experience.We point out that a
child at birth has cognitive response but without having developed the $%
\left\langle s\right\rangle $ average or $\mathcal{I}-$ consciousness until
later. A state of unconsciousness or of sleep is a ground state, precisely
the state where cognitive response to the exterior world is zero\ but the
self or $\mathcal{I}$ remains perfectly well defined and in repose.The $%
non-zero$ $\left\langle s\right\rangle $ average is the result of perfect
phase coherence of the coherent information field in the brain with a fixed
phase angle $\theta $ which represents a symmetry breaking transition
(establishment of subjectivity with respect to an objective world).
Excitation \ from this phase coherent ground state of the information field
is shown to constitute our consciousness. We also point out the underlying
structure of the dynamic memory matrix in terms of time correlation of these
self-operators.

**B.K.Chakraverty is a former director of Laboratoire de Transition de
Phase, C.N.R.S and has been a research staff member of the Solid State
Theory Group of C.N.R.S, Grenoble. He has been Emeritus director of
research, at Centre National de Recherche Scientifique, Grenoble, France.

e-mail: benoy\_chakraborty@hotmail.fr
\end{abstract}

\section{Introduction}

Physics have come a long way since the days of Newton and Galileo when it
was mainly devoted to investigations of celestial bodies. Today there is
virtually no frontier that is forbidden to the methodology of physical
investigation; from stock-market to big bang passing thorough metereology
and subatomic particles, physics tries to bring an unifying framework to the
investigating mind. The mystic of brain since time immemorial, the
difficulty of doing experiment in vivo, the belief that mind and brain have
nothing to do with each other had prevented progress in the field until very
recently. There had been in the past several classes of approach to the
brain-phenomena. There has been work revolving around the theory of neural
networks and dynamical systems \cite{hopfield} \cite{amit} \cite{haken}$.$
These approaches have the congenital difficulty of never giving emergence of
the higher brain functions or consciusness like phenomenon. Then there had
been conjectures that brain is quantum. This goes as far back as Bohr, and
as recently as R.Penrose \cite{bohr} \cite{rpenrose} \cite{satinover} \
There also has been suggested mechanisms for these quantum aspects .\cite%
{frohlich} \cite{beck} \cite{vitiello}$.$ The approach in this paper is
distinctly different. For the very first time one is introducing the concept
of $Self$ as a quantum operator reflecting genetic identity and through the
operation of this operator we have created a quantum information field.
Certain parts of this paper has echoes of Quantum decision theories \cite%
{yukalov}

In the last twenty years or so there has been an explosion of sophisticated
experimental techniques like Magnetic Resonance Imaging (MRI), Functional
Magnetic Resonance Imaging (FMRI), Positron Emission Spectroscopy (PET),
Near-Infra-red Spectroscopy (NIRS), Electroencephalography (EEG),
Magnetoencephalography (MEG) along with Computerised Tomography \& variety
of Multi-modal Imaging to track diseases of the brain but also study neural
anatomy as well as its response to a variety of stimuli. \ One can now study
some of the brain activity in-situ as well as in real time (FMRI can produce
four images every second. The brain takes half a second to to be conscious
of stimuli). Since the pioneering activity of the noted brain surgeon Wilder
Penfield\cite{penfield} who introduced electrodes into the brain to chart
out the somato-sensory map of the cortex and elicited memory pattern by
proper excitation of neural region, neurologists today are capable of
pinpointing their electrode on one single neuron and observe what happens.

Today we know a great deal about neurons, the primary agent that carry
signals to and fro between world outside and our brain inside.The human
brain is estimated to have about a hundred billion nerve cells or neurons,
two million miles of axons that take the signal down to its near neighbors
and a million billion synapses, the switch that connects one neuron with
another \cite{timgreen}. Knowledge about the physiology and the architecture
of dendrites, neurons and their axons with its synapse has developed
enormously over a century \cite{markbear}. We now know that behind every
single set of information, feeling, sensation, thought or action, a set of
neurons are involved and that there is no reason to live in the twilight of
Cartesian duality \cite{descartes} of relegating brain in some physical
space and mind elsewhere in some mental space. It would be simpler to assume
that both space is contained in the same Hilbert space where reality is
played out whether it is all measurable or not and that all of which goes on
in the brain is negotiated by the incessant flickering of these myriads of
neurons, some of them firing in unison, in a pattern with perfect inner
coherence. Their populations as well as their connections are evolutive,
never static, always adapting, developing according to ebb and flow of
information from the outside world as well as to the needs of the living
self. Everything that we do, whether experiencing an event or an emotion as
we listen to Ravi Shankar or Beethoven, our thought whether sublime or
murderous, our imagination, our desire, our acting out our will, every
single thing that becomes fabric of our mind is so because of this neural
network that subtends the mental space, that Sherrington had named the
`enchanted loom'\cite{sherrington}. We assume that there is no little `man'
or a homunculus sitting in a corner of the brain, pulling the strings of
some Cartesian theater.The Hilbert space \cite{vonneumann} where quantum
mechanics acts out, is also the mental space of brain. In this space new
quantum operators will be defined and asked to operate in perfect accordance
with the laws of causality and of thermodynamics.

On 15th january 2008, a monkey standing on a treadmill in a U.S Neurology
laboratorywith electrodes planted in some of her motor neurons made history
by making a robot stationed in Tokyo move its legs \ by the simple transfer
of the energy of her thoughts \cite{monkey}. The day is not far when
paralytics will be able to control artificial hands and legs through their
thoughts alone.Our central assumption is based on the simple belief that
thoughts and emotions carry energy and as such physics of consciousness can
be constructed from first principles.

We consider that actions of the mind can be formulated through quantum
mechanical formulation, with operators operating on Hilbert space which is
an extension of what we call our physical Hilbert space. We shall show why
the quantum description is appropriate here: continuous deformation of
neural medium is postulated to lead to discrete energy packets that we
identify as information in the mind.The quantum operators that we shall
introduce are\textit{\ operators of self}. We designate them by $S$ since
they create states or information like particles for cognitive functions $%
\left\{ \alpha \right\} $. It is these states we shall define as forming the
armature of the \textit{mental space}. We will show that these states are
formed by fundamental excitations or discrete information quanta that we
call infons. Infons are considered to be excitation out of mind field. These
excitations are taken to be boson like because a great number of them can be
imagined to be packed into a given function. These are taken to be
indistinguishible particles. This fundamental indistinguishibility seperates
quantum mechanics from classical mechanics; the classical particles move in
distinguishable space-time orbits which can be tracked continuously while
this is not true of quantum objects.From mental space we go to neural space
and assume that neurons vehicle these excitations, that they can be
exchanged from one neuron to another. Only when these boson-like information
packets develop a coherent macroscopic character by organising themselves
into distinct states or functions $\left\{ \alpha \right\} $, that we become
aware of them as distinguishable entities, as joy or pain or as good or bad.
We can use the analogy with electrons; they are indistinguihable particles.
But the way they organise, as they go from hydrogen atom to\ Uranium,
forming distinguishable orbits that each atom becomes different from its
constituting electrons and eventually completely different, from each other
giving us the infinitely rich periodic table of elements.

Our objective is to generate a global macroscopic coherent state of
information for the brain by repeated application of these $S$ operators
using a neuron \ or assembly of neurons to organise these function states.
We show that a macroscopic global coherent state of the cognitive space will
emerge. This coherent state is the eigenfunction of the global $S$ operator
and whose eigenvalue is brain's cognitive order parameter. The resultant
phase coherence is key to the whole smooth cortical synchrony or symphony.

In the next section we develop the phenomenology of the coherent brain
state. In section $2$, we present the coherent state for a single cognitive
function $\alpha $ and go on to form a global coherent state out of a
bouquet of functions. To do so, we use the coherent state formalism, due to
Glauber, so called Glauber state\cite{glauber}.

We replace the `real brain' by an organised neuron network of neurons in the
cortex communicating with each other through their synaptic connections. We
take a model brain, containing a lattice of \textit{synaptic sites in the
cortex} connected with each other through axon terminals that gather its
input through a mesh of dendrites. This is a far cry from the highly complex
human brain that has evolved over several hundred thousand years since the
Homo sapiens. We show how such an assembly has phase coherence naturally
built in and that stays in man all his life. It is at this stage that the
global $S$ operator develops a macroscopic value and a non-zero average
value $\left\langle S\right\rangle .$\textit{\ The central idea of this
paper, consists in identifying this \ operator average as our quintessential
self.} An internal executor emerges in our mind, the $\text{%
\textquotedblleft }\mathcal{I}\text{\textquotedblright }$\ that most people
say they feel exists inside their head!

In the subsection $2.2$ we will write down the thermodynamic arguments of
the emergence of this $\left\langle S\right\rangle $ $average$ and
associated spontaneous symmetry breaking.

In the section $3$, \textit{we introduce the novel idea that what we call
consciousness is nothing but a cognitive response of the neural brain to the
world.} This response function or cognitive susceptibility will be defined
in terms of these operators and applied to different states of the brain.
Section $4$ will discuss some of these results.

We may summarise this introduction by reiterating that our objective is not
whether physics can solve some of the problems of human brain (it probably
can't, like anybody else !) but whether it will allow us to think reasonably
about some of these problems.

The noted eighteenth century French physician \ Pierre Cabannis once said
that `Brain secretes thought as liver secretes bile' \cite{cabannis}. This
is almost true. Actually the function of the brain is to create
representation of the world out of the flood of incoming electro-chemical
signals that neurons vehicle; these signals are basically all alike yet
their representations in our mind are indescribable in their infinite
richness and variety.

\section{Quantum Information Field}

Mental space is taken to be a quantum information field and we suggest that
a normal functioning brain is a coherent state of this field. Why quantum
and not classical?

We ought to precise what kind of a quantum particle are we considering an
information to be-- electron like or photon like ? This is a legitimate
question since the classical and quantum limits of these two elementary
particles are slghtly different. Electromagnetic theory of Maxwell derived
its analogy from classical fluid motion. Clasical electromagnetic theory
works because in a classical light beam millions of photons are involved
where photon occupation is a continuous variable. One did not worry about
discrete nature of this number, neither did one know that a single photon
existed.\ Hence in this limit quantum theory or corpuscular description was
not needed and wave description was adequate. As far as electron went, in
the beginning, it was just the very opposite.The electron was just a
particle and like any other particle had a mass living at some point in
space and time with a definite velocity or momentum and obeyed Newton's laws
of motion. It was perfectly classical. Its quantum wave nature was
discovered much later with Scrodinger and de Broglie and then came with it,
the Heisenberg uncertainties of not knowing simultaneously its position and
momentum. It is one of the paradox of quantum mechanics that one can hardly
describe a single photon or able to write a wave function for a single
photon. Neither can one localise a photon. It was shown very early \cite%
{wigner} that this difficulty came from the fact that there was no position
operator for a photon. As a result, a single photon's probability density or
probability amplitude, its wave function at a space point can neither be
written down or normalised to unity over the space. A quantum particle on
the contrary can be described perfectly by the Schrodinger wave function $%
\psi \left( x\right) $ and like an electron can be localised. Its
probability density given by $\int \left\vert \psi \left( x\right)
\right\vert ^{2}dx$, where $\psi \left( x\right) $ is the probability
amplitude to find the particle at a space point $x$, is perfectly
normalisable and is a conserved quantity.

The choice we have made is to take information as a discrete particle like
object rather than like a photon. We consider that in a normal human brain
only an infinitesimally small amont of neural space is occupied by these
particles and as a result classical wave description like that of a light
beam is inappropriate. On the other hand why do we think that these
particles are quantum like than classical? One of the principal
characteristic of a classical particle that it can be prepared precisely at
a space point $x_{o}$ with a precise momentum $p_{o}\left( p_{o}=m\upsilon
\right) $. The limits of precision can be as fine as we want and is $\Delta
x_{o}\Delta p_{o}=0.$ This is basically because $x$\& $p$ are independent
quantities for a classical particle and we can vary one without varying the
other. We can measure one without disturbing the other and we can measure
both at the same time. All this is not true of a quantum particle. The two
quanties $x$ and $p$ are not indepenedent for a quantum object, they are
conjugate. They do not commute which means that these two quantities cannot
be measured simultaneously and if one measures $x$ one disturbs $p$ and vice
versa$.$This brings an uncertainty in the measurement given bythe famous
Heisenberg's relationship 
\begin{equation*}
\Delta x_{o}\Delta p_{o}=\hbar
\end{equation*}

Here $\hbar $ is the Planck's constant. For a classical particle if we
prepared it at the point $x_{o}$ it would remain there eternally unless
acted upon by external force which is Newton's equation of motion. For a
quantum particle on the other hand, if we did the same thing and we insist
on the particle being immobile at $x_{o}$ it will not do so. There will be
one or two things: either we will find the particle at $x_{o}$ but we will
find it at wild values of $p_{o}$ or we will find it with a momentum $p_{o}$
but its location will be anywhere in the space, with a probability given by
the Schrodinger wave function $\psi \left( x\right) .$ A classical baby in
the cradle will remain in the cradle while a quantum baby will not remain
localised but will ooze away, much to the consternation of the mother (but
since $\int \left\vert \psi \left( x\right) \right\vert ^{2}dx=1)$.$,$
mother is bound to find her baby) !

There is a fundamental reason for this quantum behaviour. In classical
physics particle motion is deterministic, determined by laws of Newton,
governing a paricle's position and its velocity. These laws are explained by
Hamilton's principle which says that trajectory a particle will choose is
determined by the principle of least action; of all the possible paths a
particle may take, the most probable one is the one that costs least action.

\begin{figure}[htb]
\begin{center}
\leavevmode
\includegraphics[scale=0.3]{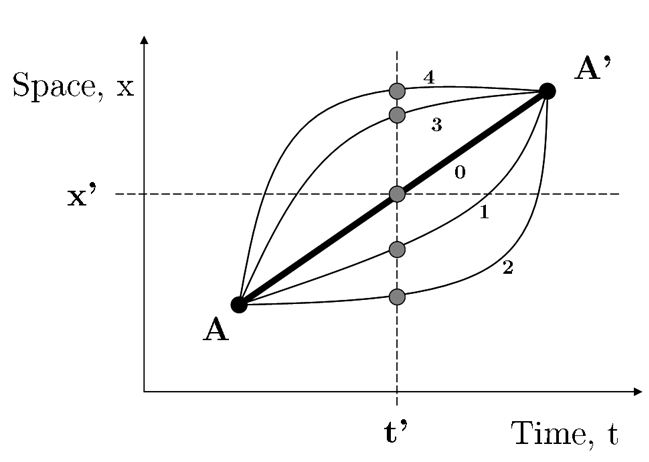}
\end{center}
\caption{Classical and Quantum Paths of an Information Particle in the Neuron}
\label{fig:classical}
\end{figure}

This least action path is the only path that a classical particle will take
(path $0$ of $A-A^{\prime }$ figure 1$)$. If we consider the action path as
a possible program, a classical information particle will execute the same
program again and again. By its very nature our mind and consequently our
brain is supposed to have free will. This means that there is no guarantee
an information particle will take the least action path in order to execute
a given function. It may well choose a variety of paths of which the least
action path is just one. Many of its paths will go over higher energy hills
and sum total of these excursions constitute the Feynman path integrals \ 
\cite{feynman} .At any instant of time $t^{\prime },$ (See Figure \ref{fig:classical}) 
the information particle may well stray away from its
classical path $0$ (point $x^{\prime },t^{\prime })$ and be found on points
indicated on the trajectories $1,2,3,4$, although its field amplitude $\psi
\left( x,t\right) $ must obey $\int_{x}\left\vert \psi \left( x,t^{\prime
}\right) \right\vert ^{2}dx=1.$ Getting away from the classical path, gives
these information particles an infinite degrees of mental freedom, which is
the reason why we can suddenly change our mind in course of an action and
take a completely different path. A computer as it is today does not have
free will and is condemned to obey the programs that had been prewritten. A
computer is classical even if it borrows neo-classical algorithms for its
functioning. It is in this strict sense information particles are quantum
objects. \textit{What their paths minimise is certainly not action but
perhaps risks involved in the action and mind will choose the path of
minimum risk.}

An information particle is created at some synaptic site $i$ but it does not
remain localised there; it hops from synapse to synapse and evetually joins
other information bits to create a coherent message. It is hopeless to ask
where a specific bit of information resides; it is delocalised, it is
disincarnate, it is everywhere and nowhere. Coherent cognitive functions
they perform can be localised and are identifiable in space and time but,
not the information bits themselves.

\textit{This is why mental space is a quantum information field}. These
quantum particles can only be generated by application of some operator on
some occupation number vectors that describe the Hilbert space of the mind.
Only a quantum description will be able to capture the underlying physics.

To start with, we have the electro-chemical signals that are coming through
different sensory channels, which seem perfectly banal, varying only in its
intensity (frequency) and duration and yet each one will become a discrete
excitation or a bit of information, exactly where and how nobody knows.
Probably the transformation (transmutation or transcription) occurs at the
somatic center of each neuron from where it will go out towards other
neurons through its axon as an action potential eventually to its
synapse.The scenario of "information" generation in the brain may be
following. Neural medium in the brain reacts to the changing
electro-chemical potential of its surrounding neurons whenever it is
disturbed by the outside world. This disturbance generates a wave like
oscillation pattern in the medium that the mind perceives as a sensation or
"information" coming in. \textit{A plane monochromatic distortion wave can
carry no information}; this is equally true of a monochromatic light beam
that cannot transport any signal,unless it is frequency or amplitude
modulated. If the "\textit{meaning}" vector of a distortion wave is taken as
amplitude of some perfectly sinusoidal wave pattern, then evidently, summed
over a few oscillation, the meaning adds to zero for a monochromatic wave of
wavelength $\lambda .$ Basically an organism is being bombarded incessantly
with facts, whose sum total information content is zero. On the other hand,
several wave lengths or facts may be called upon to interfere
constructively, so that in a small space of extension $\Delta x,$ a large
local amplitude ("information") will develop if the spread in wave number $%
\Delta k$ $\left( k=\frac{2\pi }{\lambda }\right) $ is sufficiently large.
Out of the babel of noise or constant chattering of neurons, a discrete
information bit emerges.The "information" bit ( it is a minimum uncertainty
condition because brain likes to minimise uncertainty, whose unit is $\hbar $%
) is considered to be a discrete quantum object that \textit{we have called
an infon. }From a background of a very agitated noisy neural medium, one
information quantum detaches itself almost by accident, a quantum particle
that organism finds suddenly very precious to posses. Meaning out of a
random sea of facts is an evolutionary event, revolutionary also, nothing
less nothing more but this led to cognition. This may very well be an
acceptable scenario to start with.

\textbf{The }\textit{\textbf{fundamental postulate of this communication
states that mind is a pure information space, is considered to be a quantum
field and that any state vector describing mind can only be an information
vector}.} \textit{The Fock space of the mind can be described \ by 0,1,2,...}%
$\infty $\textit{\ bits of information living in Fock or occupation number
states }$\left\{ m\right\} $ where $m=1,2,3..\infty .$ \textit{These
information quanta we shall call infons in analogy to electron or phonon or
photon}.We shall use the Dirac bra,$\left\langle A\right\vert $ or ket $%
\left\vert A\right\rangle $ notations to designate Hilbert space vectors for
example the vector $A$ \cite{Dirac}.

The infons are considered to be identical. This is so if and only if they
are excitations of the same underlying field. The often asked question "why
all electrons are identical " arises from mistakenly regarding individual
electrons as fundamental objects, when in fact it is the underlying electron
field that is fundamental. The same is true of infon particles with respect
to the underlying field which we call our mind. Quantum mechanics, in its
most general formulation, is a theory of abstract operators (observables)
acting on an abstract state space ( Hilbert space), where the observables
represent physically observable quantities and the state space represents
the possible states of our system. Each observable can be taken as a
possible degree of freedom. A classical field contains only a limited number
of degrees of freedom ( a classical electromagnetic field has only two,
local electrical and magnetic field vectors). A quantum field has unlimited,
possibly infinite degres of freedom. For our cognitive system, the
observables are the different cognitive functions, in principle there are an
unlimited number of them. We shall work in grand canonical ensemble, where
number of these fundamental excitations $\left\{ m\right\} $ will be allowed
to vary. These give rise to different functions in the mental space as
quantum superpositions of various Fock numbers $\left\{ m\right\} $; they
form the function states $\left\{ \alpha \right\} $ that live and that the
organism conserves$.$ Each function state is an eigenstate meant to preserve
the required brain function through one's whole life. The highest energy
functional state is the cognition whose ground state representation we shall
now construct. Phase coherence between infon particles which in turn gives
rise to inter-functional coherence is a result of constructive interference
between infon particles. This would not happen if these were classical
particles which never interfere and where each go their own way. It is in
this inner sense that mental space can be considered as a quantum field.

We can describe the mental field in two ways. Either in terms of mental
state wave functions in the Fock space $\left\vert \Psi _{\alpha
}\right\rangle $ where $\left\{ \alpha \right\} $ is the label of a whole
collection of states that are expected to be grouped into distinct cognitive
functions $\left\{ \alpha ,\beta ...\right\} $ as defined below. Or we go
out of the Fock space and define the mental space in terms of \ neuronal
wave functions $\left\vert i\right\rangle $ where $\left\{ i\right\} $ is
the label of a collections of cortical synapse sites $i$. Each group of some 
$\left\{ i\right\} $ is presumed to be responsible for some particular
function $\alpha .$ We insist on the cortical location \ of these synapses
which we consider to be the seat of Cognition and eventual phenomenon of
consciousness. Eventually a cognitive program or \textit{engram} \ will
emerge which is an information code in the synapse. The information that
brain generates is useful only if it is associated with some \textit{program}%
,$\left\{ p_{\alpha }\right\} .$

In the mental functional space without referring to neurons, \textit{the
shortest program one can conceive of is zero information state }$\left\vert
0\right\rangle .$\textit{\ }No information would be comprehensible without
presence of this state. The space between two written words or the silence
between two musical notes or the empty space between two strokes of colour
makes all the difference between meaning and meaninglessness. Next must be a
single information bit containing just one quantum $\left\vert
m=1\right\rangle $ ; this is like the letter $A$ or $I$ of the English
alphabet, comprising the two shortest words of the language. In general we
need a string of infons $\left\{ m\right\} $ to compose a program, strung
together in some coherent order, for it to make a sense. To get an idea of
what we call a program, let us distribute $m$ number of infons ( where $m$
goes from $0$ to $N$ ) over the different cognitive functions $\left\{
\alpha \right\} $ where $\alpha $ goes from $1$ to $M$. If we assume that
there is no restriction of number of infons that can reside in any single
function $\alpha $, then the number of distinct ways or complexions we can
arrange the $m_{\alpha }$ infons amongst $p_{\alpha }$ programs is given by
the Bose-distribution \cite{Bose}%
\begin{equation}
p\left( m_{\alpha }\right) =\frac{\left( m_{\alpha }+p_{\alpha }-1\right) !}{%
m_{\alpha }!\left( p_{\alpha }-1\right) !}  \label{bose distribution}
\end{equation}%
As an example, if we take the visual function, where we need to have
programs to, see the colours of the object, its location in space, the
different contrasts of light intensity for a given object, speed and
direction of its motion, to name at random just a few. The actual act of
seeing must integrate all these sub-functions rapidly with minimum
uncertainty. We need a whole set of programs \ covering all energy channels,
to execute a \textit{function }$\alpha .$ We can define a coherent cognitive
function $\alpha $, \textit{through sets of programs from each energy
channel }$p\left( m_{\alpha }\right) $\textit{\ as} 
\begin{equation}
\left\vert \Psi _{\alpha }\right\rangle =\sum_{m_{\alpha }=o,1,..}^{\infty
}a_{m}\left( \alpha \right) \left\vert m_{\alpha }\right\rangle
\label{definition of function}
\end{equation}

The probability amplitude $a_{m}\left( \alpha \right) $ which is a complex
number is the weight of each state $\left\vert m_{\alpha }\right\rangle $ in
the cognitive function $\alpha $ \ and is given by 
\begin{equation*}
\left\vert a_{m}\left( \alpha \right) \right\vert ^{2}=\frac{p_{m}^{\alpha }%
}{p_{m}}
\end{equation*}%
We have 
\begin{eqnarray*}
N &=&\sum_{\alpha }m_{\alpha } \\
\sum_{\alpha }p_{m}^{\alpha } &=&p_{m}
\end{eqnarray*}%
The $a_{m}^{\alpha }$ is a string of information bits $m$ that we need for
some function ; a single information bit carries no meaning. The information
content of $\alpha ,\beta ,\gamma $ etc are the different cognitive genetic
codes (different from the ones involved in the autonomous nervous functions
like respiration, heart rhythm control etc that do not depend on the
cortex). Many of these essential autonomous functions are like deep quantum
levels for the information bits and resemble orbital core states. Those
automatic nervous functions are stationary energy states like molecular
orbital states. In these states the information current is going round and
around as in electron orbitals in an atom, without dissipation, lasting a
life time; they form our daily automatism. These functions constitute sturdy
energy levels that normal outside events do not easily perturb, unless some
violent events occur.

Cognitive functions belong in this hierarchy of energy states to the highest
energy level $\epsilon _{c}$ that the infons can occupy. Continuing the
upward chain of functional hierarchy, several cognitive functions (vision,
smell, sound etc) bunch together coherently to perform a \textit{task. }Two
tasks can mutually interfere just as two classical light beams through two
slits, showing the double nature of infons: particles and wave just as light
does in the two beam experiment. The interference shows up as the difficulty
often encountered to be attentive to two cerebral tasks at the same time.

System will choose a certain set of complexions to constitute the required
function which could be vision or taste or memory or feeling. All of them
will constitute the ground or equlibrium state of the mind. We will show
later on that consciousness is a property of excited state of the cognitive
system. In the ground state one has no consciousness. We cannot have any
idea what functions will emerge in the brain of a dynosaur or a shrimp, so
numerous are the possible programs or the synaptic complexions. We shall
never know what it feels to be a bat ! Hence the question " What is it like
for bats to sense objects by echo-location ?" must remain unanswered \cite%
{Negel} The functional aspect of the program is tied to the distribution of
infons around the synaptic sites. If certain synaptic sites are never
occupied the program will wither away. And if certain function is rarely
performed, the given synaptic connection may dissolve altogether; one or
some of the terms of $p_{m}$ will not contribute. This may happen within the
life time of the individual. New functions can emerge as part of the
learning process or over a longer period, like the function of writing that
did not exist until several thousand years ago. The information space that
constitutes the Hilbert space of the mind is a functional space and is
inoperative without the neurons. This Function space constitutes the
Cognitive quantum field that will be used to construct a coherent brain
state.

\subsection{Self Operators and Coherent Brain State}

Let us introduce non-hermitian operators, that we have christened \textit{%
self operators. }It carries the instruction to fabricate information like
particles in mental space, responsible for \textit{information field and our
mental life. Self }is the expression of our genetic identity that affirms
wherever and whenever it is needed, the uniqueness of the individual. The
self operator $\mathit{s}_{\alpha }$ and its Hermitian conjugate $\mathit{s}%
_{\alpha }^{\dagger }$ has the property of destroying one infon or creating
an infon in the function state $\left\vert 1_{\alpha }\right\rangle $%
respectively \textit{out of the mental vacum} $\left\vert 0\right\rangle .$%
This is formally written as the operation%
\begin{eqnarray}
\mathit{s}_{\alpha }^{\dagger }\left\vert 0\right\rangle &=&\left\vert
1_{\alpha }\right\rangle  \label{infon operators} \\
\mathit{s}_{\alpha }\left\vert 1_{\alpha }\right\rangle &=&\left\vert
0\right\rangle  \notag
\end{eqnarray}

\bigskip Repeated application of the infon operators will generate all the
vectors of $\left\{ \alpha \right\} $ such as 
\begin{eqnarray*}
\mathit{s}_{\alpha }^{\dagger }\left\vert 1_{\alpha }\right\rangle &=&\sqrt{1%
}\left\vert 2_{\alpha }\right\rangle \\
\mathit{s}_{\alpha }\left\vert 2_{\alpha }\right\rangle &=&\sqrt{2}%
\left\vert 1_{\alpha }\right\rangle
\end{eqnarray*}

This is standard boson operator algebra; the operators are known as ladder
operators since they increase or decrease the occupation number of a state
vector by just one every time they are applied on a ket vector.Any standard
quantum mechanics text book can be consulted for details. From this
fundamental basic operators defining operations involving infons in the full
mental Hilbert space, we can go on and define operators in the Function
space $\left\{ \alpha \right\} $ $\left( \text{which is a truncated Hilbert
space}\right) ,$ through the operation%
\begin{equation}
\left\vert \Psi _{\alpha }\right\rangle =\sum_{m}a_{m}\left( \alpha \right)
\left\vert m\right\rangle _{\alpha }  \label{Function space operator}
\end{equation}

The states $\left\{ m\right\} $ are the independent linear orthogonal
vectors of Fock space defining the Hilbert space of mind while the
orthogonal linear vectors $\left\vert m_{\alpha }\right\rangle $ constitute
the mental subspace of cognitive functions.The operators $\left\{ s_{\alpha
}\right\} $ are initiators of cognitive functions and are instruction
protocols like all operators in quantum mechanics. Neurons are the conduits
of mental action, not the otherway around. Cognitive functional space $%
\left\{ \alpha \right\} $is to neurons what cyber space is to the electronic
hardware comprising a computer.

The ensemble of representations $\left\{ \left\vert \Psi _{\alpha
}\right\rangle \right\} $ constitutes the abstract space on which our whole
mental life will be constructed. All of the cognitive functions are real,
hence they belong to a Hilbert space.All reality, that which is measurable
(factual, Hermitian) and that which is non-measurable (but no less real ),
like pain or pleasure (emotional, non-Hermitian) comes out from this space.
These fundamental self operators (there are exactly three of them) are
considered to constitute the back bone of cognition system and have the
following properties :

\begin{enumerate}
\item Besides the creation operator $s_{\alpha }^{\dagger },$we have its
conjugate twin, the corresponding destruction operator $s_{\alpha }$ that
has the instruction to destroy an existing infon in the function $\alpha $,
thereby decreasing number of infons already existing in the state $%
\left\vert \Psi _{\alpha }\right\rangle $ by one. By definition vacuum state
itself is annihilated by its action $s_{\alpha }\left\vert 0\right\rangle =0$%
, for all $\alpha .$

\item The combined action of these two operators\ is the third operator,
called preservation or number operator and helps count the total number of
infons in a given neuron, when it operates on that state.%
\begin{equation*}
s_{\alpha }^{\dagger }s_{\alpha }=n_{\alpha }
\end{equation*}%
To emphasise the operator character of the number operator $n$ we have 
\begin{equation}
n_{\alpha }\left\vert i\right\rangle =p_{\alpha }\left\vert \alpha
\right\rangle  \label{number op}
\end{equation}%
This operation or measurement gives us the total number of information like
particles in the function $\alpha $. The creation and the destruction
operators are taken to be non-hermitian while $n_{\alpha }$ is hermitian.%
\newline

\item We consider the infons as Bose particles. In contrast to Fermions
which occupy a spot in space, only one at a time (in the absence of spin),
Bosons have the advantage that they can be generated at any space as many as
one wants by repeated application of the creation operator on vacuum. Any
number $p_{\alpha }$ of \textit{infons} can crowd into any single function $%
\left\vert \Psi _{\alpha }\right\rangle $.

\item The different functions $\left\{ \alpha \right\} $ commute. Translated
into simple language, it means that any number of cognitive functions can be
measured (felt) simultaneously.We write this as Bose commutation
relationships \cite{walecka}%
\begin{eqnarray}
\left[ s_{\alpha },s_{\beta }\right] &=&\left[ s_{\alpha }^{\dagger
},s_{\beta }^{\dagger }\right] =0 \\
s_{\alpha }^{\dagger }\left( x\right) ,s_{\beta }\left( x^{\prime }\right)
&=&\delta _{\alpha \beta }\delta _{xx^{\prime }}
\end{eqnarray}
\end{enumerate}

The first set of relationships tell us that those pair of operators commute
at equal time and that their actions are simultaneously measurable. Because
they commute, they are independent and do not interfere with each other. We
will show below that this equal time commutability proceeds from the fact
that the set of operators $\left\{ s_{\alpha }\right\} $ generate their
individual eigen values \ when they operate \textit{on the same coherent
wave function}. The second set of relationships imply that any two operators 
$\left\{ s_{\alpha }^{\dagger }\left( x\right) ,s_{\beta }\left( x^{\prime
}\right) \right\} $ are orthogonal .

To get a coherent wave function signifying the coherent brain state, let us
first focus on just one\ single cognitive function $\alpha $. The defining
function state for $\left\vert \Psi _{\alpha }\right\rangle $ shows that a
varying population of infons $m$ is needed for each function $\alpha .$ The
coherent Glauber state of infons \ is written as the wave function $%
\left\vert \Psi _{\alpha }\right\rangle $, as \cite{wolf} 
\begin{equation}
\left\vert \Psi _{\alpha }\right\rangle =\sum_{m=0,1,2,...}a_{m}\left(
\alpha \right) \left\vert m\right\rangle _{\alpha }  \label{glauber p-state}
\end{equation}%
The significant aspect of this wave function is the possibility that at any
given time there can be any number of infons in the function $\alpha $; the $%
a_{m}\left( \alpha \right) $\ are complex coefficients. If we choose these
coefficients judiciously, then the different probability amplitudes $%
a_{m}\left( \alpha \right) $ of each of the Fock state $\left\vert
m\right\rangle $ will add up constructively to give a macroscopic amplitude
of infons only if they have a common phase angle $\theta _{\alpha }.$ When
this happens we shall get the coherent state $\left\vert \Psi _{\alpha
}\right\rangle $.

An exactly equivalent formulation of the Glauber state, can be given
explicitly in terms of $s_{\alpha }^{\dagger }$. The coherent state, in the
zeroth order is given by the exponential operation 
\begin{equation}
\left\vert \Psi _{\alpha }\right\rangle =\exp \left( \phi _{\alpha
}s_{\alpha }^{\dagger }\right) \left\vert 0\right\rangle
\label{Zerothorder state}
\end{equation}

Here although it is not visible yet, the exprssion has the parameter $\phi
_{\alpha }$ which will turn out to be the hidden cognitive order of the
function $\alpha .$ This state has the expansion%
\begin{equation}
\left\vert \Psi _{\alpha }\right\rangle =\left\vert 0\right\rangle +\phi
_{\alpha }s_{\alpha }^{\dagger }\left\vert 0\right\rangle +\frac{\left( \phi
_{\alpha }s_{\alpha }^{\dagger }\right) ^{2}}{2!}\left\vert 0\right\rangle
+...  \label{Psi-expansion}
\end{equation}%
This expansion shows that the coherent state is made out of varying number
of infons. To understand the coherent state, we may write the operator
expression%
\begin{equation}
s_{\alpha }(\phi _{\alpha })=\exp ^{-\phi _{\alpha }s_{\alpha }^{\dagger
}}s_{\alpha }\text{ }\exp ^{\phi _{\alpha }s_{\alpha }^{\dagger }}=s_{\alpha
}+\phi _{\alpha }  \label{translation operator}
\end{equation}

We see that the action of the exponential operator is to translate the
destruction operator by a complex number $\phi _{\alpha }.$ This gives the
key property of the coherent state as being the eigenstate of the
destruction operator%
\begin{equation}
s_{\alpha }\left\vert \Psi _{\alpha }\right\rangle =\exp ^{\phi _{\alpha
}s_{\alpha }^{\dagger }}\exp ^{-\phi _{\alpha }s_{\alpha }^{\dagger
}}s_{\alpha }\text{ }\exp ^{\phi _{\alpha }s_{\alpha }^{\dagger }}\left\vert
0\right\rangle =\phi _{\alpha }\left\vert \Psi _{\alpha }\right\rangle
\label{eigenstate of the derstruction op}
\end{equation}

This result follows when we use the fact $s_{\alpha }\left\vert
0\right\rangle =0.$ The result also shows that $\phi _{\alpha }$ is the
eigenvalue of the destruction operator$.$ Since $s_{\alpha }$ is a
non-hermitian operator, the eigenvalue can only be complex. This result
points out that the operator average $\left\langle s_{\alpha }\right\rangle $
is precisely $\phi _{\alpha }.$%
\begin{equation}
\left\langle \Psi _{\alpha }\right\vert s_{\alpha }\left\vert \Psi _{\alpha
}\right\rangle =\phi _{\alpha }  \label{individual neuron order}
\end{equation}%
That the complex parameter $\phi _{\alpha }$ is in reality an order
parameter can be seen from 
\begin{equation*}
\left\langle \Psi _{\alpha }\right\vert n_{\alpha }\left\vert \Psi _{\alpha
}\right\rangle =\left\langle \Psi _{\alpha }\right\vert s_{\alpha }^{\dagger
}s_{\alpha }\left\vert \Psi _{\alpha }\right\rangle =\phi _{\alpha }^{\ast
}\phi _{\alpha }=\left\langle N_{\alpha }\right\rangle
\end{equation*}%
Since $\left\langle N_{\alpha }\right\rangle $ is just a number, number of
infons on an average involved in the function $\alpha ,$ we can write down
the order parameter as a complex scalar quantity%
\begin{equation*}
\phi _{\alpha }=\sqrt{\left\langle N_{\alpha }\right\rangle }\exp i\theta
_{\alpha }
\end{equation*}

The different functions $\left\{ \Psi _{\alpha }\right\} $ \textit{are
distinguishable}. Although the electrical signals coming through the neurons
are all alike to start with , what ultimately distinguish them one from the
other is the response they provoke in the different sensory channels.The
global coherent state due to all $\left\{ \Psi _{\alpha }\right\} $
functions can now be written down. For all these functions we can write for
the global coherent cognitive wave function, the product wave function 
\begin{equation}
\left\vert \Psi _{C}\right\rangle =\Pi _{\alpha }\left\vert \Psi _{\alpha
}\right\rangle  \label{cognitive wave function}
\end{equation}

\ \ This can be expanded as 
\begin{equation}
\left\vert \Psi _{C}\right\rangle =\Pi _{\alpha }\exp \left( \phi _{\alpha
}s_{\alpha }^{\dagger }\right) \left\vert 0\right\rangle =\exp
^{\sum_{\alpha }\left( \phi _{\alpha }s_{\alpha }^{\dagger }\right)
}\left\vert 0\right\rangle =\exp ^{S^{\dagger }\Phi _{C}}\left\vert
0\right\rangle  \label{global neuron coherent state}
\end{equation}

We have written $\Phi _{C}$ as a \ $\upsilon \times 1$ column matrix 
\begin{equation}
\Phi _{C}=\left( 
\begin{array}{c}
\phi _{\alpha } \\ 
\phi _{\beta } \\ 
.. \\ 
\phi _{\upsilon }%
\end{array}%
\right)  \label{Column & row matrix}
\end{equation}%
We also write global creation operator $S^{\dagger }$ as a $1\times \upsilon 
$ row matrix%
\begin{equation*}
S^{\dagger }=\left( s_{\alpha }^{\dagger }....s_{\upsilon }^{\dagger }\right)
\end{equation*}%
Then we have%
\begin{equation*}
S^{\dagger }\Phi _{C}=\sum_{\alpha ,...\upsilon }s_{\alpha }^{\dagger }\phi
_{\alpha }
\end{equation*}%
This allowed us to write as we did the global coherent state 
\begin{equation*}
\left\vert \Psi _{C}\right\rangle =\exp ^{S^{\dagger }\Phi _{C}}\left\vert
0\right\rangle
\end{equation*}%
The coherent state $\left\vert \Psi _{C}\right\rangle $ has the nice
property of being able to single out a given function order parameter when
it is acted upon by the function field operator $s_{\alpha }.$ 
\begin{equation*}
s_{\alpha }\left\vert \Psi _{C}\right\rangle =\phi _{\alpha }\left\vert \Psi
_{C}\right\rangle
\end{equation*}

The global cognitive wave function $\left\vert \Psi _{C}\right\rangle $
allows simultaneous measurements in all functional channels $\alpha $ and
this is why these operators $\left\{ s_{\alpha }\right\} $ commute.

we can write for the global order parameter $\Phi _{C}$%
\begin{equation}
\Phi _{C}^{\ast }\Phi _{C}=\frac{1}{M}\sum_{\alpha }\Phi _{\alpha }^{\ast }%
\text{ }\Phi _{\alpha }=\frac{1}{M}\sum_{i}\left\langle N_{\alpha
}\right\rangle =\frac{N_{C}}{M}  \label{global population}
\end{equation}

Here $N_{C}$ is the global average information population in $the$ $\
cortical$ $brain$ \ summed over all the cognitive functions, $M$. Now a
global cognitive order parameter $\Phi _{C}$ has emerged with one single
phase $\theta _{C}$ to signify over-all phase coherence of the information
field. Expression of equation \ref{global population} allows us to write for
the global cognitive order in the form%
\begin{equation}
\Phi _{C}=\sqrt{\frac{N_{C}}{M}}\text{ }\exp i\text{ }\theta _{C}
\label{global phase and order}
\end{equation}

There are several key points we would like to make at this stage:

(a) To obtain the global order, we have summed over all neuron labels. This
emphasises the fact that the cognitive order parameter $\Phi _{C}$
represents in reality the full mental landscape. The individual label and
phase of each neuron has disappeared from the global cognitive order which
has emerged with its own global phase $\theta _{C}$ $indepenent$ $of$ $space$
$and$ $time$.

(b) The global order parameter can be defined as the operator average of the
global destruction operator of self $S$ which we write as%
\begin{equation}
\Phi _{C}=\left\langle \Psi _{C}\right\vert S\left\vert \Psi
_{C}\right\rangle  \label{nonzero S-average}
\end{equation}%
Here $S$ is the column matrix \ representing destruction operators%
\begin{equation*}
S=\left( 
\begin{array}{c}
s_{\alpha } \\ 
s_{\beta } \\ 
... \\ 
s_{\upsilon }%
\end{array}%
\right)
\end{equation*}%
Since $S$ is one of the three matrix elements of \textit{Self, we take the
bold step to call this order parameter $\mathcal{I}$. }We make the
identification 
\begin{equation}
\Phi _{C}=\mathcal{I}  \label{phi as I}
\end{equation}

Now the self operator has taken a macroscopic significance. "$\mathcal{I}$%
\textbf{\ am" has emerged as a result of global phase coherence between N}$%
_{C}$\textbf{\ information bits. The meaning of the global cognitive order
is $\mathcal{I}$. This phase coherence is brought about by more and more
rapid information transfer through synaptic connections between neurons. A
critical neuron band-width or \ connectivity must occur before, $\mathcal{I}$
can emerge.}

(c) The unique global phase angle $\theta _{C}$ for $\Phi _{C}$ with which
order parameter emerges is a symmetry breaking transition. Any other $\theta 
$ would have been equally good from the point of view of total energy of the
cognitive system, but this $\theta _{C}$ and only this one, the order
parameter $\Phi _{C}$ of the brain system has chosen and retains throughout
one's whole life. We have named this $\mathcal{I,}$ precisely because this
unique $\theta _{C}$ confers on each individual his individuality, the
imprint of an unique personality. The subjective \textit{self }given by $%
\left\langle S\right\rangle $ breaks the symmetry of the mental space $%
\left\{ m\right\} $; a subjective -objective symmetry so to speak. From this
point onwards, \textit{self }and \textit{self}$-consciousness$ emerge as the
hallmark of a stable personality.

(d) Mathematically the unique global phase $\theta _{C}$, translates the
fact that the infon population $N_{C}$ is a variable number and the coherent
brain ground state $\left\vert F_{C}\left( \theta \right) \right\rangle $\
that fixes $\theta $ can be expressed in the form%
\begin{eqnarray}
\left\vert \Psi _{C}\left( \theta \right) \right\rangle &=&\sum_{N_{C}}\Psi
\left( N_{C}\right) \exp iN_{C}\theta _{C}  \label{uncertainty reletion} \\
\Delta N_{C}\Delta \theta _{C} &\sim &1
\end{eqnarray}

\textit{The uncertainty relationship between phase locking of the global
wave function and its information content is fundamental to the coherency of
all brain processes.We must allow this number to fluctuate if we are to have
a macroscopic coherent cognitive state.}

\subsection{Synaptic Self and Spontaneous Symmetry Breaking \ }

In the preceding section we have constructed a globally coherent cognitive
state $\left\vert \Psi _{C}\right\rangle $ associated with the cognitive
order parameter $\Phi _{C}$ that we have called the first person $\mathcal{I}
$. We have worked entirely in the mental landscape defined by its diverse
cognitive functions. Everything was done as if outside world did not exist.
But developing coherent cognitive functions in the absence of interaction
with outside world is as useless as developing an alphabet or language that
no one would use. As a matter of fact, one suspects that the cognitive
functions that would survive are precisely those that help us to cope with
the world in a Darwinian sense.

The world connects to the mind through our neurons. Mind also expresses
itself through the same neuron network. Thus the neuron is the go-between
mind and world, a window for the mind within and for the world without.
Neurons connect with other neurons through the synaptic sites. While just
before a baby is born, neurons are being created at the astonishing rate of
250,000 neurons per minute, right after birth synaptic connections between
those neurons are being made at the astronomical rate of several million
connections per second! \cite{ledoux}. One can make a strong case that
synaptic connections are essential for information transfer betwwen
different regions of brain and that synaptic sites may well be where infons
are stored. At least this is the view we shall adopt.

We shall introduce field operators of self $\psi _{i}$ connected with info
creation at synaptic sites,$i$. The corresponding creation operator is $\psi
_{i}^{\dagger }$ which when operates on the vacuum state creates one infon
in the state vector $\left\vert 1_{i}\right\rangle .$We write 
\begin{equation}
\psi _{i}^{\dagger }\left\vert 0\right\rangle =\left\vert 1_{i}\right\rangle
\label{site operator}
\end{equation}

The local field operators $\psi _{i}$ can be wriiten in terms of the
internal function space basis operators $s_{\alpha }$%
\begin{equation}
\psi _{i}=\sum_{\alpha }\phi _{\alpha }\left( i\right) s_{\alpha }
\label{localfield function of mentalstate}
\end{equation}

And similarly for $\psi _{i}^{\dagger }.$ Here $\phi _{\alpha }\left(
i\right) $ is a complex probability amplitude of finding the projection of
the mental state $\alpha $ on the synaptic site $i.$ We can obtain the
average value of the synaptic site operator $\left\langle \psi
_{i}\right\rangle $ by using the cognitive wave function \ref{cognitive wave
function} 
\begin{equation}
\left\langle \psi _{i}\right\rangle =\left\langle \Psi _{c}\right\vert \psi
_{i}\left\vert \Psi _{c}\right\rangle  \label{average synaptic operator}
\end{equation}

To illustrate, suppose we have a $\Psi _{c}$composed of just two cognitive
functions $\alpha $ and $\beta .$

\bigskip Then we have for the wave function $\Psi _{c}=\left( 
\begin{array}{c}
\Psi _{\alpha } \\ 
\Psi _{\beta }%
\end{array}%
\right) $ 
\begin{equation*}
\left\langle \psi _{i}\right\rangle =\left[ \Psi _{\alpha }^{\ast }\text{ }%
\Psi _{\beta }^{\ast }\right] \left[ 
\begin{array}{c}
\psi _{i}^{\alpha \alpha }\text{ }\psi _{i}^{\alpha \beta } \\ 
\psi _{i}^{\beta \alpha }\text{ }\psi _{i}^{\beta \beta }%
\end{array}%
\right] \left[ 
\begin{array}{c}
\Psi _{\alpha } \\ 
\Psi _{\beta }%
\end{array}%
\right]
\end{equation*}
\ Here $\psi _{i}^{\alpha \alpha }$ $is$ , $\left\langle \Psi _{\alpha
}\right\vert \psi _{i}\left\vert \Psi _{\alpha }\right\rangle $ and
similarly for the other matrix elements.

This is rewritten as 
\begin{equation*}
\left\langle \psi _{i}\right\rangle =\left[ 
\begin{array}{c}
\left\vert \Psi _{\alpha }\right\vert ^{2}\text{ \ }\Psi _{\alpha }\Psi
_{\beta }^{\ast } \\ 
\Psi _{\beta }\Psi _{\alpha }^{\ast }\text{ \ }\left\vert \Psi _{\beta
}\right\vert ^{2}%
\end{array}%
\right] \left[ 
\begin{array}{c}
\psi _{i}^{\alpha \alpha }\text{ }\psi _{i}^{\alpha \beta } \\ 
\psi _{i}^{\beta \alpha }\text{ }\psi _{i}^{\beta \beta }%
\end{array}%
\right]
\end{equation*}

This can be also written as 
\begin{equation*}
\left\langle \psi _{i}\right\rangle =trace\text{ }\left( \rho \psi
_{i}\right)
\end{equation*}%
The infon density matrix $\rho $ (which is a $M\times M$ square matrix ) has
the usual definition 
\begin{equation*}
\rho =\left\vert \Psi _{c}\right\rangle \left\langle \Psi _{c}\right\vert = 
\left[ 
\begin{array}{c}
\Psi _{\alpha } \\ 
\Psi _{\beta } \\ 
. \\ 
etc%
\end{array}%
\right] \left[ \Psi _{\alpha }\text{ }\Psi _{\beta }...etc\right]
\end{equation*}

\bigskip Now we are in a position to define global synaptic self average $%
\Phi _{s}$ as 
\begin{equation}
\Phi _{s}=\left[ 
\begin{array}{c}
\left\langle \psi _{i}\right\rangle \\ 
\left\langle \psi _{k}\right\rangle \\ 
\left\langle \psi _{l}\right\rangle \\ 
etc%
\end{array}%
\right]  \label{synaptic self average}
\end{equation}

\textit{We impose global synaptic self average to be the same as the
cognitive functional average and equate both to }$I$\textit{. We write }$I$%
\textit{\ }$=\Phi _{c}=\Phi _{s}$\textit{. This implies }%
\begin{equation*}
\left\vert \Phi _{c}\right\vert ^{2}=\left\vert \Phi _{s}\right\vert
^{2}=\sum_{i}\left\vert \left\langle \psi _{i}\right\rangle \right\vert
^{2}=\sum_{i}n_{i}=N_{c}
\end{equation*}

The statement made just above is capital. \textit{It says what goes in the
mind goes in the synapses; there is no way to distinguish our cognitive self
as epitomised by }$I$\textit{\ from our synaptic self}. \cite{ledoux}.

As the order parameter develops in the ground state, long range correlation
develops between local order between different synaptic sites , say $i$ $\&$ 
$j.$ If the distance between $i$ $\&$ $j$ goes to $\infty $ but correlation $%
\left\langle \psi \left( i\right) \psi ^{\ast }\left( j\right) \right\rangle 
$ remains finite, then we have a genuine Bose condensation \cite{Bose} in
human brain. Because of the finite dimension of our system this is
impossible to have. A less restrictive condition of having something like a
bose-condenstate is to rewrite this correlation in an alternate form. We
write 
\begin{eqnarray}
\left\langle \psi \left( i\right) \psi ^{\ast }\left( j\right) \right\rangle
&=&\left\langle \sum_{\alpha }\phi _{\alpha }\left( i\right) s_{\alpha
}\sum_{\beta }\phi _{\beta }^{\ast }\left( j\right) s_{\beta }^{\dagger
}\right\rangle  \label{offdiaogonal correlation} \\
&=&\sum_{\alpha ,\beta }s_{\alpha }s_{\beta }^{\dagger }\left\langle \phi
_{\alpha }\left( i\right) \phi _{\beta }^{\ast }\left( j\right) \right\rangle
\notag \\
&=&\sum_{\alpha }N_{\alpha }\left\langle \phi _{\alpha }\left( i\right) \phi
_{\alpha }^{\ast }\left( j\right) \right\rangle  \notag
\end{eqnarray}

Here we have used the commutation properties of the operators $s_{\alpha }$ ,%
$s_{\beta }^{\dagger }$ etc introduced in the last section. This is
off-diagonal information correlation between two different synaptic sites
and can have a macroscopic value if the condensate density $N_{\alpha }$
develops in one of the functional channels.This is closest we shall get to a
bose-condensed state in these inhomogeneous finite size systems.

The coherent cognitive state is a symmetry broken state, as we explained in
the last section. Let us be a little more specific.

There is a whole general class of systems that show spontaneous symmetry
breaking in their ground state while their dynamics (hamiltonian) is
invariant of that symmetry\cite{coleman}.The ferromagnet is a familiar
example: its global magnetisation chooses to lie in an arbitrary direction,
while it could have chosen any other direction without any extra energy
cost. Superfluid He or Superconductors are other examples from condensed
matter physics, where the order parameter chooses a global unique phase
while its free energy does not depend on that phase angle. In our case the
cognitive order parameter does the same although we do not know the exact
nature of the hamiltonian $H_{0}$that we need to describe the dynamics or
evolution of the information field. What we need to note is that the
coherent cognitive wave function $\left\vert \Psi _{C}\right\rangle $ and
its associated order parameter $\Phi _{c}$ were constructed by repeated
action of the self operator on the vacuum state $\left\vert 0\right\rangle $%
. This $\left\vert 0\right\rangle $ is nothing but the bare inherited
genetic magma \ from the very instant that the child was conceived. There
was no reference to the world as yet. We need to confront this order
parameter to the world which the new born baby will face. It is convenient
to introduce the world as some external perurbation $H^{^{\prime }}$to see
if the unique ground state $\left\vert \Psi _{C}\right\rangle $ engendered
by $H_{0}$ remains intact in the absence of the perturbation as we go to the
limit of no world. This is extremely relevant since everyday we go to this
limit when we fall asleep and every time we do so we need to recover the
same unique ground state $\left\vert \Psi _{c}\right\rangle $ with the
cognitive order $\mathcal{I}$ .

Let us write the total hamiltonian governing the cognitive brain as 
\begin{equation}
H=H_{0}+H^{\prime }  \label{total hamiltonian}
\end{equation}

The perturbation due to world (this includes interaction with one's own
body) is written as%
\begin{equation}
H^{\prime }=\sum_{i}\eta _{i}\left[ \psi _{i}\Omega _{i}^{\ast }+\psi
_{i}^{\dagger }\Omega _{i}\right] =\sum_{i}\eta _{i}H_{i}^{\prime }
\label{world perturbation}
\end{equation}

Here the world designated by $\Omega _{i}$ acts at the synaptic site $i$
locally with the operator $\ \ \psi _{i}$through some suitable coupling
constant $\eta _{i}.$We assume that the main part of the Hamiltonian $H_{0}$
had done its job in creating the unperurbed ground state $\left\vert \Psi
_{C}\right\rangle $, with a corresponding ground state energy $E_{C},$ which
is the lowest energy of the coherent cognitive state, of the brain at
repose. Due to coupling $\eta ,$ both the ground state wave function as well
as the state energy will be shifted to $\left\vert \Psi _{C}\left( \eta
\right) \right\rangle $ and $E_{C}\left( \eta \right) .$Let us call the new
total hamiltonian by $H(\eta )$ and write 
\begin{equation}
E_{C}\left( \eta \right) =\left\langle \Psi _{C}\left( \eta \right)
\left\vert H(\eta )\right\vert \Psi _{C}\left( \eta \right) \right\rangle
\label{perturbed energy}
\end{equation}

The change in ground state energy can be written as ( due to a trick first
used by Pauli)

the so called coupling constant integration 
\begin{equation}
\Delta E_{C}=E_{C}\left( 1\right) -E_{C}\left( 0\right)
=\sum_{i}\int_{0}^{1}d\eta _{i}\left\langle \Psi _{C}\left( \eta _{i}\right)
\left\vert H_{i}^{\prime }\right\vert \Psi _{C}\left( \eta _{i}\right)
\right\rangle  \label{change in energy}
\end{equation}

\bigskip This perturbation generates a new operator average $\left\langle
\psi _{i}\right\rangle $ that we can write as 
\begin{equation}
\left\langle \psi _{i}\right\rangle =\frac{\partial \left( \Delta
E_{C}\right) }{\partial \Omega _{i}}=\int_{0}^{1}d\eta _{i}\left\langle \Psi
_{C}\left( \eta \right) \left\vert \frac{\partial H_{i}^{\prime }}{\partial
\Omega _{i}}\right\vert \Psi _{C}\left( \eta \right) \right\rangle
\label{operator average}
\end{equation}%
We do this integration by seperating it into two parts as%
\begin{equation*}
\left\langle \psi _{i}\right\rangle =\left\langle \Psi _{C}\left( 0\right)
\left\vert \frac{\partial H_{i}^{\prime }}{\partial \Omega _{i}}\right\vert
\Psi _{C}\left( 0\right) \right\rangle +\int_{\eta \neq 0}^{1}d\eta
_{i}\left\langle \Psi _{C}\left( \eta \right) \left\vert \frac{\partial
H_{i}^{\prime }}{\partial \Omega _{i}}\right\vert \Psi _{C}\left( \eta
\right) \right\rangle
\end{equation*}

If the first term of the right hand side survives even in the absence of
coupling to the world then we have a symmetry broken ground state given by
the local cognitive order parameter average $\left\langle \psi
_{i}\right\rangle _{0}.$ We can then write 
\begin{equation*}
\left\langle \psi _{i}\right\rangle =\left\langle \psi _{i}\right\rangle
_{0}+\delta \left\langle \psi _{i}\right\rangle _{0}
\end{equation*}%
The sources or the `world' $\Omega $ \& $\Omega ^{\ast }$ were introduced \
in order to select a unique equilibrium state--so as to set the `alignement$%
^{\text{'}}$ of the cognitive system just as a magnetic field does for the
ferromagnet. These sources induce non-vanishing values of the field
operators,$\langle \psi _{i}^{\dagger }\rangle _{0}$ and $\left\langle \psi
_{i}\right\rangle _{0}.$ For a normal system that does not show spontaneous
symmetry breaking, these field expectation values vanish when the sources
are turned off. But in a symmetry broken state, this does not occur. The non
zero operator averages remain intact even when there is no external
source.This \ result shows us that $\Phi $ has the broken symmetry : $\Phi
\rightarrow \Phi _{o}$ even when the world $\Omega \rightarrow 0.$\textit{%
Translated into more mundane cognitive terms, this says that as we fall
asleep, the world }$\Omega \rightarrow 0$\textit{\ but the cognitive order
parameter, }$\mathcal{I}$\textit{\ returns to the base value, characterising
the equlibrium ground state. World is lost during sleep, but not }$\mathcal{I%
}$\textit{.}

$N_{o}$ and $\theta $ are conjugate quantities. This is exemplified by the
uncertainty relationship 
\begin{equation*}
\triangle N_{o}\triangle \theta \succeq 1
\end{equation*}%
The simple reason that phase and particle number are conjugate quantities
imply that their simultaneous measurement is limited by the Heisenberg
uncertainty principle. Consequently, boson like particles can either be in
an eigenstate of particle number or of phase. The eigenstate of particle
number means a system with fixed population of infons, and is a localised
state or a neuron with no connection to other neurons. This phase can be
called $a-state.$The second state of the information system, named $b-state$
is the one where the information is fluid but the phase coherence has very
short range in space and time.This is a mixed state, neither localised nor
completely fluid, at best is an incoherent mixture of the two and is not an
eigenstate. The eigenstate of phase is a superfluid. This is the state where
information particles live and move coherently from synapse to synapse.This
state we will call the $state$ $c.$ We can characterise each of these states
by 
\begin{eqnarray*}
\left\langle \exp i\theta _{i}\right\rangle &=&0,state\text{ }a \\
\left\langle \exp i\theta _{i}\right\rangle &\neq &0,\text{ }\left\langle
\exp i\left( \theta _{i}-\theta _{j}\right) \right\rangle =0\text{ ; }state%
\text{ }b \\
\left\langle \exp i\left( \theta _{i}-\theta _{j}\right) \right\rangle &\neq
&0,i-j\rightarrow \infty ,state\text{ }c
\end{eqnarray*}%
Here $i$ and $j$ are neuron positions.

We can think of $\alpha -state$ as belonging to worms or single cellular
creatures possessing a few or no neurons to speak of. $b-state$ can be
expected to belong to babies, less than 2 yrs old and higher domestic
animals, creatures that are perfectly conscious but not of themselves ;there
is, as yet no long range phase coherence, conscious experience is there but
is fragmented. There may be a ghost of $\mathcal{I}$ but it is more like the
smile of a cheshire cat ! In the $c-state$, long range phase correlation
between neurons are firmly established and brain has entered its coherent
state; that of adult human brain. Penrose \cite{penrose} had posed the
question whether a one cellular creature like a paramecium or a bacterium
(which does not even have a neuron) can have consciousness? Our answer seems
to be quite unambiguous-- it cannot. It lives in the state $\alpha $ (neuron
or no neuron ), no order parameter can form locally ; even if it did, phase
and amplitude fluctuation will kill all coherence as it invariably does in
one dimensional systems. A 2-yr old baby posseses already an $\mathcal{I}$
and recognises himself in the mirror. From $\mathcal{I}$=0 at birth, the
individual has gone to the free energy minimum of $\mathcal{I}\neq 0$, all
due to the tremendous explosion in synaptic connectivity in those first two
years after a child's birth. These three states mimic closely condensed
phase of bosonic systems, namely, localised insulting state, disordered
boson glass phase \& symmetry broken superfluid phase , $a,b$ \& $c$ phases
respectively \cite{philipps}.

\subsection{Thermodynamics of Cognitive Order}

To make some of these ideas more quantitative, we express the Hamiltonian \ $%
H$ of $M-neurons$, in two parts, a part which is internal to the brain
system, $H_{o}$ and a part that brings about perturbation due to interaction
of the infons with the world, $H^{\prime }.$We consider that in the absence
of the world, the cognitive system develops the coherent order parameter $%
\Phi _{C},$ called now $\Phi _{C}^{o}$ to indicate that it is the
unperturbed ground state, engendered by $H^{o}.$

Let us write the partition function to obtain the relevant thermodynamic
quantities to obtain \ the equlibrium order parameter as an extremum of
Hemholtz free energy and see how this shifts in the presence of $H^{\prime }$%
.We introduce real time $t$ but this symbol can also be replaced by
imaginary time if we have to.We write the perturbibation due to external
world as

\begin{equation*}
H^{\prime }=\int dt\int dr_{i}\left[ \psi _{i}^{\dagger }(t)\Omega
(i,t)+\psi _{i}(t)\Omega (i,t\right]
\end{equation*}

We also have used 
\begin{equation*}
\psi (i,t)=\exp ^{iH_{o}t}\psi _{i}(o)\exp ^{-iH_{o}t}
\end{equation*}

Here $H_{o}$ is the unperturbed original hamiltonian that we have not
specified so far. We can write for the grand partition function through
functional integration%
\begin{equation}
Z\left[ \Omega ,\Omega ^{\ast }\right] =\int \left[ d\psi _{i}\right] \left[
d\psi _{i}^{\ast }\right] \exp \left\{ \int dtL+\int dt\int dr\left[ 
\begin{array}{c}
\psi _{i}^{\ast }(t)\Omega (r,t) \\ 
+\psi _{i}(t)\Omega ^{\ast }(r,t)%
\end{array}%
\right] \right\}  \label{partition function}
\end{equation}%
Here $L$ is the Lagrangian given by%
\begin{equation}
L=H_{o}-i\int dr_{i}\text{ }\psi _{i}^{\ast }(t)\frac{\partial }{\partial t}%
\psi _{i}(t)  \label{lagrangian}
\end{equation}%
Here we have replaced the operators $\psi $ and $\psi ^{\dagger }$ by
functional integration variable $\psi $ and $\psi ^{\ast }.$ In this
formulation the self operators are integrated away and the partition
function is expressed only in terms of the world. The effective action or
Helmholtz free energy is given by 
\begin{equation}
F\left[ \Omega ,\Omega ^{\ast }\right] =\ln Z\left[ \Omega ,\Omega ^{\ast }%
\right]  \label{helmholtz potential}
\end{equation}

By simple differentiaion we get%
\begin{eqnarray}
\frac{\partial F\left[ \Omega ,\Omega ^{\ast }\right] }{\partial \Omega (i,t)%
} &=&\frac{1}{Z}\frac{dZ}{d\Omega (i,t)}=\left\langle \psi _{i}^{\dagger
}(t)\right\rangle  \label{first derivative of partition fn} \\
\frac{\partial F\left[ \Omega ,\Omega ^{\ast }\right] }{\partial \Omega
^{\ast }(i,t)} &=&\frac{1}{Z}\frac{dZ}{d\Omega ^{\ast }(i,t)}=\left\langle
\psi _{i}(t)\right\rangle  \notag
\end{eqnarray}

To see this clearly, it is convenient to consider the expectation values of $%
\langle \psi _{i}^{\dagger }(t)\rangle $ and $\left\langle \psi
_{i}(t)\right\rangle $ as the independent variables (rather than the sources 
$\Omega $ \& $\Omega ^{\ast }$) by carrying out a functional Lagrange
transformation which defines the Gibb's potential $\Lambda $%
\begin{equation}
\Lambda =\int dt\int dr_{i}\left[ \psi _{i}^{\ast }(t)\Omega (i,t)+\psi
_{i}(t)\Omega ^{\ast }(i,t\right] -F\left[ \Omega ,\Omega ^{\ast }\right]
\label{Gibbs potential}
\end{equation}

\bigskip Consider variation of this equation with respect to $\Omega $ \& $%
\Omega ^{\dagger }$. We obtain%
\begin{equation*}
\partial \Lambda =\int dt\int dr_{i}\left\{ \partial \langle \psi
_{i}^{\dagger }\rangle \Omega +\Omega ^{\ast }\partial \left\langle \psi
_{i}\right\rangle \right\}
\end{equation*}%
Rest of the terms give zero. Thus we may regard $\Lambda $ as a functional
of $\left\langle \psi \right\rangle $ and $\left\langle \psi ^{\dagger
}\right\rangle .$ This gives us 
\begin{eqnarray}
\frac{\partial \Lambda \left[ \left\langle \psi \right\rangle ,\left\langle
\psi ^{\dagger }\right\rangle \right] }{\partial \left\langle \psi ^{\dagger
}(i,t\right\rangle } &=&\Omega (i,t)  \label{definition of external world} \\
\frac{\partial \Lambda \left[ \left\langle \psi \right\rangle ,\left\langle
\psi ^{\dagger }\right\rangle \right] }{\partial \left\langle \psi
(i,t\right\rangle } &=&\Omega ^{\ast }(i,t)  \notag
\end{eqnarray}

These derivatives give us a functional definition of external world
parametrised by $\Omega .$ We can go to an external world which is constant
in space and time, $\Omega (i,t)\rightarrow \Omega $ and $\Omega ^{\ast
}(i,t)\rightarrow \Omega ^{\ast }.$ In this case the expectation values of
the field operators must also be constant. Now, we can write the extensive
Gibb's potential in the intensive form 
\begin{equation}
\Lambda =\beta VG(\left\langle \psi _{i}\right\rangle ,\langle \psi
_{i}^{\dagger }\rangle )  \label{gibbs free energy}
\end{equation}

Here $G$ is Gibb's free energy per unit volume and $V$ is the volume of the
system. Let us a global cognitive field order parameter as 
\begin{equation*}
\left\langle \psi _{i}\right\rangle =\left\langle \Psi \right\rangle =\Phi
\end{equation*}%
In the limit of sources uniform locally over each neuron, we have 
\begin{eqnarray}
\frac{\partial G(\left\vert \left\langle \Psi \right\rangle \right\vert ^{2})%
}{\partial \left\langle \psi _{i}\right\rangle } &=&\Omega _{i}^{\ast }
\label{fixing of external source} \\
\frac{\partial G(\left\vert \left\langle \Psi \right\rangle \right\vert ^{2})%
}{\partial \langle \psi _{i}^{\dagger }\rangle } &=&\Omega _{i}  \notag
\end{eqnarray}

The correct thermodynamic state is determined as a stationary point of the
effective potential or Gibb's free energy. At the equilibrium point $%
\left\langle \psi _{i}\right\rangle =\left\langle \Psi \right\rangle _{eq}$
we must have

\begin{eqnarray}
\left( \frac{\partial G(\left\vert \left\langle \Psi \right\rangle
\right\vert ^{2})}{\partial \left\langle \psi _{i}\right\rangle }\right)
_{\left\langle \psi _{i}\right\rangle =\left\langle \Psi \right\rangle
_{eq}} &=&o \\
\left( \frac{\partial G(\left\vert \left\langle \Psi \right\rangle
\right\vert ^{2})}{\partial \left\langle \psi _{i}^{\dagger }\right\rangle }%
\right) _{_{\left\langle \psi _{i}^{\dagger }\right\rangle =\left\langle
\Psi ^{\dagger }\right\rangle _{eq}}} &=&o  \notag
\end{eqnarray}%
This expression is true at all extremum. This shows us immediately that at
the minimum of the Gibb's free energy, which will determine the cognitive
order parameter with it's unique symmetry broken state, the world $\Omega $
vanishes. When we remember from the last section that $\left\langle \Psi
_{c}\right\rangle =\Phi _{s}=\mathcal{I}$, then we can draw the conclusion 
\textit{that when we are in the equlibrium state of the cognitive sysem, the
world vanishes. In full anaesthesia or in sleep }$\mathcal{I}$ \textit{%
remains perfectly intact and every time we wake up we do retrieve our }$%
\mathcal{I}$. This is the unambiguous demonstration that the brain lives in
a spontaneous symmetry broken state, akin to many condensed state systems
including superfluid.

\begin{figure}[htb]
\begin{center}
\leavevmode
\includegraphics[scale=0.4]{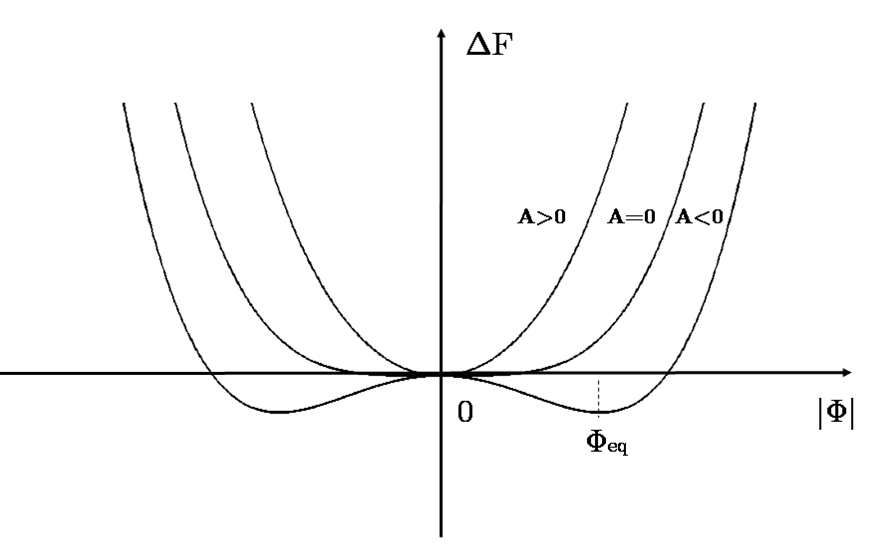}
\end{center}
\caption {Free energy against amplitude of order parameter : $\mathbf{A}\succ \mathbf{0}$, babies $\preceq $ 2 yrs; $\mathbf{A=0,}$ threshold of self; $\mathbf{A\mathbf{\prec }0,}$babies $\succ 2$ yrs, self $\mathcal{I}$ formed}
\label{fig:cognitiveorder}
\end{figure}

From now onwards, we shall call the mental space containing the cognitive
order parameter $\left\langle \Psi _{c}\right\rangle ,$ as an $\mathcal{I}%
-field.$ The ground state of this field occurs where $\left\langle \Psi
_{c}\right\rangle $ gives a free energy minimum. From a semiclassical point
of view ,$\left\langle \Psi _{c}\right\rangle $ can be considered as a field
which interacts with itself through a potential ( also called the Gibb's
free energy function, $G(\left\vert \left\langle \Psi _{c}\right\rangle
\right\vert ^{2})$ written in the Ginzburg-Landau form\cite{degennes}%
\begin{equation}
G(\left\vert \left\langle \Psi \right\rangle \right\vert ^{2})=\mathbf{A}%
\left\vert \Phi \right\vert ^{2}+\mathbf{B}\left\vert \Phi \right\vert ^{4}
\end{equation}%
Here $\left\vert \left\langle \Psi \right\rangle \right\vert $ $=\Phi $, is
the ampltude of the average of the order parameter which we have shown to be
a complex quantity with an amplitude and a phase. Let us start from the
non-symmetry broken phase , with a positive value of the parameter $\mathbf{A%
}$ that gives the minimum at $\left\vert \Phi \right\vert =0.$ Eventually
when the phase transition to a $\left\vert \Phi \right\vert \neq 0$ phase
would occur, symmetry breaking will take place, fixing the phase once for
all. This expression for Gibb's free energy will assure us a minimum of the
free energy at $\Phi =\Phi _{eq}$if the parameter $\mathbf{A}$ is $\prec 0$
and $if$ the coefficient of the fourth order term $\mathbf{B}$ is $\succ 0.$
The nature of these curves is shownin Figure \ref{fig:cognitiveorder}.

It is clear that the vital parameter that brings about this minimum is when $%
\mathbf{A}$ changes sign from positive value where the minimum of cognitive
order is zero to a negative value,when some non-zero value $\Phi _{eq}$
develps, which is the value at equilibrium given by $\left( \frac{\partial G%
}{\partial \left\vert \Phi \right\vert }\right) _{\left\vert \Phi
\right\vert =\left\vert \Phi \right\vert _{eq}}=0.$ This ground state of the 
$\mathcal{I}-field$ occurs at%
\begin{equation}
\left\langle \Psi \right\rangle _{eq}=\sqrt{\frac{-\mathbf{A}}{\mathbf{B}}}%
\exp -i\theta
\end{equation}%
This ground state is infinitely degenerate in $\theta $, lying as it does at
the bottom of the so called Mexican hat potential defined by
\textquotedblleft a ring of minima\textquotedblright\ for whatever be the
value of $\theta .$ If this phase angle $\theta $ can be arbitrarily chosen
at each point in space and time, then the interaction of $\left\langle \Psi
\right\rangle _{eq}$ with external world would move it continually as if $%
\left\langle \Psi \right\rangle $ were a free particle. But it is not ; it
is a coherent state. The ground state of the system is required to be
unique, so that phase must be fixed once for all, at all points of space and
time. Thus symmetry breaking is self imposed to rid \textit{self }of the
tyranny of the outside world !

We know that a baby as born has no sense of self as yet and does not know
who he or she is until at certain age $\sim 2$ years old. We also know that
as as soon as a child is born, there is an explosion in his brain of
synaptic connections, at the astonishing rate of $\sim $ two
millions/second, which is a measure of density of information flow from one
neuron to its neighbors via its axon terminal ( there are about $10^{4}$
synaptic connections per neuron ). The vital parameter $\mathbf{A}$ is
connected to this synaptic connectivity ($Appendix).$ Beyond a critical
value of synaptic connectivity $\mathbf{A}$ becomes negative and
all-important cognitive self can emerge as a coherent order parameter $%
\left\langle \Psi \right\rangle _{eq}$ or $\mathcal{I}$. We will sketch in
the appendix a possible scenario of how it comes about.

The fluctuation out of the Mexican hat potential well and is governed by the
coefficient of the second order term and gives\textbf{\ }$\mathbf{A}$%
\begin{equation}
\left( \mathbf{A}\right) _{\left\langle \psi \right\rangle =\Phi _{eq}}=%
\frac{1}{2}\left( \frac{\partial ^{2}G(\Phi ^{2})}{\partial ^{2}\left\vert
\Phi \right\vert }\right) _{\left\langle \psi \right\rangle =\Phi _{eq}}
\end{equation}

\section{Cognitive Response}

\subsection{Cognitive Response and Consciousness}

The problem of Consciousness is considered by many modern philosophers as
the "hard problem " \cite{shear}. The Well known Australian philosopher
Chalmers\cite{chalmers} details in a book why it is so hard and also which
ones are easy problems; these include an objective study of the brain. In a
more modest answer to some of these issues, that avoids erudite pitfalls, it
is meaningful to \textit{define consciousness as part of cognitive response
of the brain to the world.} Having defined the ground state of the cognitive
system as the $\mathcal{I-}field$, it seems sensible to ask what the excited
state is like. The excitation comes when world presents itself and interacts
with self operator. In the ground state where world is absent by
construction , there is no world to couple with ; there is no consciousness,
as a result. Consciousness is one of the function of the excited state of
the cognitive system. \textit{it is a pure Response Function}. The problem
is still hard but we have cleared a small space to work on and part of the
problem becomes more tractable .

In this simple approach, we will couple external world designated by $\Omega 
$ to the global self operator $\Psi ^{\dagger },$ where $\Omega $ is
considered to be an infinite source and sink of information. Here $\Psi $
and $\Omega $ are matrices $\left\{ \psi _{i}\right\} $ and $\left\{ \Omega
_{i}\right\} .$We define cognitive response $\chi $ as response of the brain
to perturbation $H^{\prime }$ due to external world$.$ We use linear
response theory \cite{nozieres}. and write 
\begin{equation*}
H^{\prime }(t^{\prime })=-\text{ }\eta \text{ }\Omega (t^{\prime })\Psi
^{\dagger }(t^{\prime })+h.c
\end{equation*}

Here we presume that world is turned on at time $t^{\prime }$ very
slowly,coupled to the self \ creation operator $\Psi ^{\dagger }(t^{\prime
}) $ with a coupling constant $\eta .$ For the time being we omit the
spatial index, to keep it simple. This perturbation will give the retarded
response 
\begin{equation}
\delta \left\langle \Psi (t)\right\rangle =-\frac{i}{\hslash }\int_{-\infty
}^{t}dt^{\prime }\left\langle \left[ \Psi (t),H^{\prime }(t^{\prime })\right]
\right\rangle  \label{Order param change}
\end{equation}

$\mathcal{I}$ feels the change $\delta \left\langle \Psi (t)\right\rangle $
and is conscious of the change because the first order change $\delta
\left\langle \Psi (t)\right\rangle $ brings about a second order change of
the free energy of the ground state $\sim \left( \delta \left\langle \Psi
(t)\right\rangle ^{2}\right) .$\textit{This response constitutes awareness
of }$I$\textit{\ to the world and we define it as cognitive perception.}
Only a small part of this perception is a conscious perception and we call
it our consciousness. \textit{Precisely Consciousness results from that part
of the response function which is dissipative or imaginary}. There is a
whole part of the response function that we are not conscious of. Because
cognitive response is considered to be ruled by causality, with response
lagging behind the stimulation in time, we have .retarded response function
or susceptibility given by 
\begin{equation}
\delta \left\langle \Psi (t)\right\rangle =\int_{-\infty }^{t}dt^{\prime }%
\text{ }\chi _{R}\left( t-t^{\prime }\right) \text{ }\Omega (t^{\prime })
\label{order oarameter response}
\end{equation}%
Here the susceptibility is defined by the commutator 
\begin{equation}
\chi _{R}\left( t-t^{\prime }\right) =-\frac{i}{\hslash }\theta (t-t^{\prime
})\left\langle \left[ \Psi (t),\Psi ^{\dagger }(t^{\prime })\right]
\right\rangle  \label{defn of susceptibility}
\end{equation}

The $\theta -function$ where $t\succ t^{\prime }$ \ assures the causality,
cause preceding effect.

\ \ We all know what it is to be unconscious. We also know what it is to be
conscious or waking up to the hustle and bustle of the world. Unconscious
state is a state of repose. Our brain is at its free energy minimum and
world $\Omega $ is absent at this minimum. In deep sleep or general
anesthesia, awareness of the world around us disappears. We take it for
granted that it should be so. But there is a paradox in this. At this free
energy minimum where $\mathcal{I}$ is very much present, so is the cognitive
response function,$\chi _{R}\left( \omega \right) $ which we just defined .
Then why does the awareness go away? The precise answer lies in very nature
of the cognitive response function which will also permit us to give an 
\textit{operational definition of consciousness and unconsciousness.} Here
the cognitive susceptibility is a retarded function (subscript $R$ ) given
by the operator average 
\begin{equation*}
\chi _{R}\left( t-t^{\prime }\right) =\left\langle \Psi (t)\Psi ^{\dagger
}(t^{\prime })\right\rangle ,\text{ }t\geq t^{\prime }
\end{equation*}%
Defining cognitive susceptibility as a linear response function to the
world, we have made the implicit assumtion of causality. Its fourier
transform is 
\begin{equation*}
\chi _{R}\left( \omega \right) =\int_{o}^{\infty }d(t-t^{\prime
})\left\langle \Psi (t)\Psi ^{\dagger }(t^{\prime })\right\rangle \exp \left[
i\omega \left( t-t^{\prime }\right) \right]
\end{equation*}%
The causality imposes on the $\chi _{R}\left( \omega \right) $ the
Krammer's-Kronig relationship, \textit{so that the response is a complex
quantity}, having a real and an imaginary part (the two parts are related
through Hilbert transform).%
\begin{equation}
\chi _{R}\left( \omega \right) =\chi ^{\prime }\left( \omega \right) +i\chi
"\left( \omega \right)
\end{equation}

The imaginary part of the response function $\chi "\left( \omega \right) $
monitors \textit{real neuronal excitation} from the ground state. This is
the part that would give rise to real sensations, emotion and eventual
dissipation of the excitation back into the outside world as heat and sensed
by the organism as fatigue. \textit{We define the imaginary part as
Consciousness. }Since $\chi "\left( \omega \right) $ is odd in $\omega ,\chi
"\left( \omega \right) =0$, at $\omega =0.$This explains why there is no
conscious response when brain is at the free energy minimum. This minimum is
situated at $\left\langle \Psi \right\rangle =0,$for a baby$\leq 2$ yrs old
and at $\left\langle \Psi \right\rangle =\mathcal{I}$, for all other cases
where selfhood has been achieved. We are unconscious at this precise point.
Real part of cognitive response $\chi ^{\prime }\left( \omega \right) $ is
finite of course due to virual infon excitations. Conscious perception
results only with the real excitations. \textit{Subsequent decay of real
excitations confer on them a life time or the time needed for us to be
conscious of an event; the imaginary part consequently has a spectral weight
over which the excitation energies are spread out,which we perceive as a
conscious experience}. This rainbow hue of spectral spread is sensed by self 
$\left( \text{even when }\mathcal{I}\text{ is not yet formed}\right) $ as a
direct perception of the world in all its splendour, called "qualia " of
conscious experience \cite{chalmers} .The full $\chi _{R}\left( \omega
\right) $ has poles in the lower energy $\left( \epsilon =\hbar \omega
-i\delta \right) $ plane that define the exact excitation energies with a
small imaginary part $\delta $ .

We will address this issue in both cases: for babies less than 2-yrs old
when\ one is at the free energy minimum of $\left\langle \Psi \right\rangle
=\Phi =\mathcal{I=}0$ and for children above that age when $\left\langle
\Psi \right\rangle =\Phi =\mathcal{I}\neq 0$ when one is in the symmetry
broken phase.

The approach to the coherent state free energy minimum is heralded by the
static real part of $\chi _{R}\left( \omega =0\right) $ which one does
identifiy as $inverse$ of $\mathbf{A}$ 
\begin{equation*}
\frac{1}{\chi _{R}\left( \omega =0\right) }=\left( \mathbf{A}\right) =\left( 
\frac{\partial ^{2}G(\left\vert \left\langle \Psi \right\rangle \right\vert
^{2})}{\partial ^{2}\left\vert \left\langle \Psi \right\rangle \right\vert }%
\right)
\end{equation*}
which goes to zero (susceptibility diverges, see appendix) as the cognitive
order begins to develop. The imaginary part related to dissipation during
cognitive perception \textit{is what we assert to be conscious}. It includes
the emotive part of the response, as the perception manifests itself,
through visible emotion, palpable sensation, rapid eye motion or increased
heart beat, skin temperature rise or sudden blips in the E.E.G signal in the 
$\gamma -$frequency region often characteristic of the awake conscious
state. The real part $\chi ^{\prime }\left( \omega \right) $ is related to
the lossy part $\chi "\left( \omega ^{\prime }\right) $through 
\begin{equation}
\chi ^{\prime }\left( \omega \right) =\frac{1}{\pi }\int_{-\infty }^{\infty
}d\omega ^{\prime }\chi "\left( \omega ^{\prime }\right) P\frac{1}{\omega
^{\prime }-\omega }  \label{kramer'skronog}
\end{equation}

\begin{figure}[htb]
\begin{center}
\leavevmode
\includegraphics[scale=0.5]{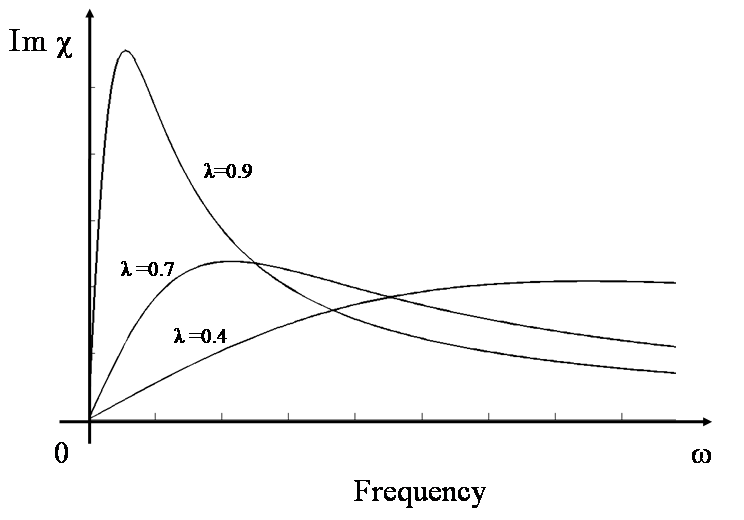}
\caption{Approach to self as synaptic connectivity increases to threshold of consciousness.}
\end{center}
\label{fig:frequency}
\end{figure}

Here $P$ is the principal value integral over the lossy part of
susceptibility. The integral says that if the real part of cognitive
susceptibility on the left is to become large at $\omega =0,$ it can be so
if the integral on the right with the imaginary part becomes more and more
intense around the low energy response. This is clearly seen if we come down
from the normal phase where there is as yet no cognitive order $\mathcal{I}$
is still=0 $\left( \text{for a baby }\prec 2\text{ yrs}\right) $and approach
the point when the real part of the cognitive response starts diverging. The
imaginary part (see Appendix ) of the susceptibility of any one given neuron 
$i$ can be written 
\begin{equation*}
\func{Im}\chi _{i}^{l}=\frac{\rho _{o}\omega \tau }{\left( 1-\lambda \rho
_{o}\right) ^{2}+\lambda ^{2}\rho _{o}^{2}\omega ^{2}\tau ^{2}}
\end{equation*}%
Here $\omega $ is excitation energy measured from the equlibrium energy
state and $\tau $ a characteristic relaxation time for relaxation of the
excitation, $\rho _{o}$is a density of states of these info particles and $%
\lambda $ a characteristic energy scale of synaptic connectivity. As the
system starts going critical at $\lambda \rho _{o}\rightarrow 1,$when the
real part starts to diverge, the imaginary part becomes more and more peaked
at low energy. In the attached Figure \ref{fig:frequency} we plot, $\func{%
Im}\chi _{i}^{l}$ showing the series of curves reflecting the intensity of
conscious experience as $\lambda \rho _{o}\rightarrow 1$ and the child $%
\left( \succ \text{2 yrs old}\right) $ acquires non-zero cognitive order $%
\left( \mathcal{I}\neq 0\right) $.

This may explain why early childhood experiences are so intense.This
abundance of low energy excitations is probably at the root of intensity of
some conscious experience, and its `qualia'.

There are several key remarks that should be made to make clear the ground
on which we stand.

(a) Although at $\left\langle \Psi \right\rangle $ at $\left\langle \Psi
\right\rangle _{eq}=\mathcal{I,}$ external world has vanished, one can be
marginally conscious of dream like phenomenon. However, if one can avoid
falling asleep and achieve this ground state through some techniques of
profound meditation, the cognitive susceptibility will consist in
consciousness of self without any awareness of the world. The pecrceiving
self $\mathcal{I}$ is very much present.

(b) The world springs into being as soon as $\left\langle \Psi \right\rangle 
$ moves out of the free energy minimum\ at $\mathcal{I}$ and positions
itself at any other point on the curve where the slope is given by 
\begin{equation*}
\left( \frac{\partial G(\left\vert \left\langle \Psi \right\rangle
\right\vert ^{2})}{\partial \left\langle \Psi \right\rangle }\right)
_{\left\langle \psi \right\rangle =\left\langle \psi ^{\prime }\right\rangle
\neq \left\langle \psi \right\rangle _{eq}}=\Omega
\end{equation*}%
The cognitive response will consist now of a significant part which is
conscious response \ that we will call consciousness defined below. This
consists in awareness of the world and of self.

(c) Two space-time events $\Omega \left( i,t\right) $ and $\Omega \left(
j,t^{\prime }\right) $ will have relationship with each other when and only
when, the events are negotiated though the cognitive susceptibility. This is
given by the free energy piece%
\begin{equation}
\Delta F=\Omega \left( i,t\right) \chi _{R}\left( i-j,t-t^{\prime }\right)
\Omega \left( j,t^{\prime }\right)  \label{connection between events}
\end{equation}

This has the immediate consequence that relationship discovered between
phenomenon is mediated by our sensorial perception and is not independent of
the cognitive mechanism that observes it. There is one little comment about
the nature of physics that this last relationship underlines. In classical
physics, observations between facts and relationships between them are out
there to be discovered. In quantum physics, observables are only those that
are not disturbed by the observation process itself; not all relationships
between observables are possible. In the physics of consciousness,
observation depends on the observer and relationships between objects are
dependent on the perception $\chi _{R}$ of the observer. An absolute
relationship between observales does not exist, as best is an illusion. But
since the operator of self $S$ is non-hermitian, its average, called $%
\mathcal{I}$ \ the observer,is not Hermitian either. As a result,it is not a
measurable or can be object of observation.

\subsection{Memory and dynamics of Perception}

We have seen that a symmetry broken ground state emerges which is immuable
in space and time, characterised by the quantity $\mathcal{I}$, the
cognitive order parameter which is a macroscopic manifestation of our
penultimate $\mathit{self}$. We are permitted to replace the original vacuum
state $\left\vert 0\right\rangle $ that we started with by the new ground
state which we call $\left\vert \mathcal{I}\right\rangle $. While $%
\left\vert 0\right\rangle $ represented the nothingness of no information
state of the original pristine mind, $\left\vert \mathcal{I}\right\rangle $ 
\textit{is the full coherent state }that the self operator $S$ or its
synaptic counterpart $\Psi $ has sculpted out of this primordial nothingness.

The basic infon propagator from one neuron $i$ to another $j$ is wriiten as
the Green's function 
\begin{equation}
g\left( i-j,t-t^{\prime }\right) =\left\langle \mathcal{0}\left\vert \psi
(i,t)\psi ^{\dagger }(j,t^{\prime })\right\vert \mathcal{0}\right\rangle
\label{infon propagator}
\end{equation}

This is formally obtained from the Free energy expression \ $F$ of the
preceding section by\ differentiating it two times (here the average is over 
$\left\vert \mathcal{I}\right\rangle $%
\begin{equation*}
\frac{\partial ^{2}F}{\partial \Omega (j,t^{\prime })\partial \Omega ^{\ast
}(i,t)}=\left\langle \psi (i,t)\psi ^{\dagger }(j,t^{\prime })\right\rangle
\end{equation*}%
This one particle Green's function constitutes the building block of our
dynamic day to day or episodic memory in contrast to the ground state memory
of the reservoir of infon particles $N_{o}$ built out of genetic material
that gave rise to $\mathcal{I}$. If we differentiate the free energy $2M$ $%
times\ $we get the $M$-point correlation function ,%
\begin{equation*}
\frac{\partial ^{2M}F}{\partial \Omega (j,t^{\prime })....\partial \Omega
(M,t^{\prime })\partial \Omega ^{\ast }(i,t)..\partial \Omega ^{\dagger
}(M,t).}=\left\langle \psi (i,t)...\psi (M,t)\psi ^{\dagger }(j,t^{\prime
})...\psi ^{\dagger }(M,t^{\prime })\right\rangle
\end{equation*}%
We make use of Bloch-deDominicis decomposition \cite{Bloch} to get all
combinations of two by two factors to get the average of a product of
creation and annihilation operators that gives us%
\begin{equation*}
\left\langle \boldsymbol{.....}\right\rangle =\sum_{all\text{ }%
i,j}\left\langle \psi (i,t)\psi ^{\dagger }(j,t^{\prime })\right\rangle
\end{equation*}%
This is our dynamic memory matrix, $M_{R}$ a $M\times M$ matrix, given by 
\begin{equation*}
\left\langle \boldsymbol{.......}\right\rangle \text{ }=\left[ 
\begin{array}{c}
\psi _{i}\left( t\right) \\ 
\psi _{j}\left( t\right) \\ 
... \\ 
\psi _{M}\left( t\right)%
\end{array}%
\right] \left[ \psi _{i}^{\dagger }\left( t^{\prime }\right) ....\psi
_{M}^{\dagger }\left( t^{\prime }\right) \right] =\mathcal{M}_{R}
\end{equation*}%
The expression $\left[ \psi _{i}^{\dagger }\left( t^{\prime }\right)
....\psi _{M}^{\dagger }\left( t^{\prime }\right) \right] $ is a short hand
for expressing that at some past time $t^{\prime }$ a page of a book was
written with infons on different synaptic sites $i,j,...M$ etc. It is like
an instantaneous photograph at the instant $t^{\prime }$ of the states of
occupation of the synapses.The ket associated just on its left the long
column $\left[ 
\begin{array}{c}
\psi _{i}\left( t\right) \\ 
\psi _{j}\left( t\right) \\ 
... \\ 
\psi _{M}\left( t\right)%
\end{array}%
\right] $ is telling us that the same page is being read at the very present
moment $t$, or another photograph of the same set of sites is being taken at
the instant, $t$.$.$ If the tensor product has a non-zero joint amplitude
i.e if the two sets of photograhs match, then one has memory of what
happened at the instant $t^{\prime }.$ If there is decoherence in
propagation of infons between these two times, then memory will be impaired.
This can be written more succintly as retaded susceptibility function $%
t\succeq t^{\prime }$, with a subscript $R$

\begin{equation}
\chi _{R}\left( t-t^{\prime }\right) =-\frac{i}{\hslash }\theta (t-t^{\prime
})\left\langle \left[ \Psi (t),\Psi ^{\dagger }(t^{\prime })\right]
\right\rangle =matrix\text{ }\mathcal{M}_{R}\text{ }  \label{Memory matrix}
\end{equation}

\bigskip Here $\Psi (t)$ is the synaptic site destruction operator matrix%
\begin{equation*}
\Psi (t)=\left[ 
\begin{array}{c}
\psi _{i}\left( t\right) \\ 
\psi _{j}\left( t\right) \\ 
... \\ 
\psi _{M}\left( t\right)%
\end{array}%
\right]
\end{equation*}%
and $\Psi ^{\dagger }(t^{\prime })$ is the creation operator matrix given by 
$\left[ \psi _{i}^{\dagger }\left( t^{\prime }\right) ....\psi _{M}^{\dagger
}\left( t^{\prime }\right) \right] .\left[ \psi _{i}^{\dagger }\left(
t^{\prime }\right) ....\psi _{M}^{\dagger }\left( t^{\prime }\right) \right]
.$ It is instructive to look at the Fourier transform $\chi _{R}\left(
\omega \right) $of $\chi _{R}\left( t-t^{\prime }\right) $ as $t-t^{\prime
}\rightarrow \infty $ \ We write 
\begin{equation*}
\chi _{R}\left( \omega \right) =\int_{\infty }^{0}d(t-t^{\prime })\exp
i\omega \left( t-t^{\prime }\right) \chi _{R}\left( t-t^{\prime }\right)
\end{equation*}%
There is no gurantee that such a Fourier transform exists, particularly if
it exists in the limit of $\chi _{R}\left( \infty \right) .$ That would
imply permanent memory. But if it does, we can write is as%
\begin{equation*}
\chi _{R}\left( \omega \right) =\chi _{R}\left( \infty \right) \delta \left(
\omega \right) +\chi _{R}\left( \omega \neq 0\right)
\end{equation*}%
The second term of this expression contains contribution of all the short
term memories, while the first term tries to catch all episodic memories
which we call our autobiography. It is static and time does not efface it
and retrievable at any instant $t$ , if we had the means to do so. They seem
to be gone most of the time but they are not. Under external stimulation
sometimes they surface bursting into our consciousness as fishes out of the
deep sea, surprising us.

\begin{figure}[htb]
\begin{center}
\leavevmode
\includegraphics[scale=0.5]{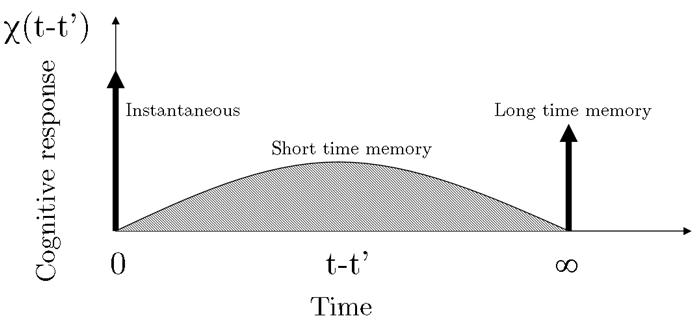}
\caption{Dynamic memory. Time correlation of cognitive response : instantenaeous, long time and short time memory.}
\end{center}
\label{fig:memory}
\end{figure}

Finally, there is the instantaneous memory given by 
\begin{equation}
\chi _{R}\left( t=t^{\prime }\right) =\int_{\omega }\chi _{R}\left( \omega
\right) d\omega  \label{instantaneous memory}
\end{equation}%
Instantaneous memory is the integrated energy response of the neural system.
These three regimes are shown in Figure \ref{fig:memory}.

The infinite temporal correlation between infon particles when it exists
becomes the fabric of our dynamic memory matrix. \textit{This memory tape is
eternally preserved except in pathological situations}. The whole aspect
related to decoherence and memory loss is intended in a future publication.

The cognitive susceptibility,is given by single particle Green's function or
propagator\ because it describes propagation of information from one
space-time point to another. Trasformed in the Fourier space, it describes
the same information carrying a momentum $q$ ( although momentum is not a
good quantum number in a non-homogeneous \ system) and excitation energy $%
\omega .$Perturbation due to external world causes an excitation from the
ground state $\left\vert I\right\rangle .$ we will designate this excitation
by the global consciousness operator $\varphi _{c}$%
\begin{equation}
\varphi _{c}=\Psi -\left\langle \Psi _{C}\right\rangle
\end{equation}

The consciousness annihilation operator has the operational definition%
\begin{equation}
\varphi _{c}\left\vert \mathcal{I}\right\rangle =0
\end{equation}%
$\varphi _{c}^{\dagger }$ creates quasiparicles. \textit{The state }$%
\left\vert \mathcal{I}\right\rangle $\textit{\ is the vacuum of
consciousness carrying quasiparticles. When one is at the ground state }$%
\left\vert \mathcal{I}\right\rangle $ one has no consciousness. We will be
in the Heisenberg representation where this symmetry broken ground state $%
\left\vert \mathcal{I}\right\rangle $ is immobile in time while the
consciousness operators are time dependent. Hermitian conjugate of the
annihilation operator $\varphi _{c}$ is the creation operator $\varphi
_{c}^{\dagger }$ of a consciousness quasiparticle given by 
\begin{equation*}
\varphi _{c}^{\dagger }\left\vert \mathcal{I}\right\rangle =1\left\vert
c\right\rangle
\end{equation*}%
Here $\left\vert c\right\rangle $ is an excited state describing
consciousness.

Quasiparticle excitation is a single particle response. We shall now outline
a microscopic sketch of what is involved in single particle excitation that
causes consciousness.We express it in terms of local consciousness operator, 
$\varphi _{c}(i,t)$ as%
\begin{equation}
\varphi _{c}(i,t)=\psi (i,t)-\left\langle \psi (i)\right\rangle
\label{fluctuation operator}
\end{equation}%
Write for single particle Green's function or the excited state information
propagator%
\begin{equation}
g\left( i-j,t-t^{\prime }\right) =\left\langle \varphi _{c}(i,t)\varphi
_{c}^{\dagger }(j,t^{\prime })\right\rangle  \label{real space green fn}
\end{equation}%
We can expand just one single particle propagator, of a piece of information
going from neuron $j$ to $i$

\begin{eqnarray}
\left\langle \varphi _{c}(i,t)\varphi _{c}^{\dagger }(j,t^{\prime
})\right\rangle &=&g\left( i-j,t-t^{\prime }\right) =g_{o}\left(
i-j,t-t^{\prime }\right)  \notag \\
&&+g_{o}\left( j-k,t^{\prime }-t"\right) \Sigma \left( k-i,t"-t\right)
g\left( i-k,t-t^{^{\prime \prime }}\right)
\end{eqnarray}%
\newline

The first term on the right hand side is the amplitude of direct propagation
of information from $j\rightarrow i$ and from $t^{\prime }\rightarrow t.$The
second term describes the same process but takes into fact that propagation
may not be direct and can go though many an indirect channels (like an
intermediate neuron $k$ )before reaching the destination neuron $i$.We can
go to the Fourier space ( since neurons are distinguishable, $q$ is a poor
quantum number),%
\begin{equation*}
\varphi _{c}\left( q,\omega \right) =\int_{t}\int_{r}d(i-j)d(t-t^{\prime
})\varphi _{c}(i-j,t-t^{\prime })\exp iq(r_{i}-r_{j})\exp i\omega
(t-t^{\prime })
\end{equation*}%
we define 
\begin{equation}
g\left( q,\omega \right) =\left\langle \varphi \left( q,\omega \right)
\varphi ^{\dagger }\left( q,\omega \right) \right\rangle
\label{Propagator in fourier space}
\end{equation}%
This propagator has poles where the amplitude becomes very large and occurs
at specific values of $\omega .$The corresponding excitation carries a $q$
and $\omega $ label as it travels. Thus every thought and emotion which
correspond to these excitations carry real momentum and real energy.%
\begin{equation}
g\left( q,\omega \right) =g_{o}\left( q,\omega \right) +g_{o}\left( q,\omega
\right) \Sigma \left( q,\omega \right) \mathcal{g}\left( q,\omega \right)
\label{fourier space green fn}
\end{equation}

\begin{equation}
g\left( q,\omega \right) =\frac{g_{o}\left( q,\omega \right) }{1-g_{o}\left(
q,\omega \right) \Sigma \left( q,\omega \right) }  \label{comact G(q,w)}
\end{equation}

The function $\Sigma \left( q,\omega \right) $is the fourier transform of
the self energy of the fluctuation green's function written earlier in real
space-time as $\Sigma \left( r"-r,t"-t\right) .$ The non-interacting green's
function is $g_{o}\left( q,\omega \right) $ is a high energy process,has a
pole at $\omega =\epsilon _{q},$ which \ is the energy needed to excite a
particle out of the condensate or ground state. It automatically confers the
same pole to $g\left( q,\omega \right) .$ 
\begin{equation*}
g_{o}\left( q,\omega \right) \approx \frac{1}{\omega -\epsilon _{q}}
\end{equation*}

\textit{This high energy excitation is subconscious \ perception process
because the information is carried swiftly from one spot to another. This is
the amplitude mode that is expected to have an energy gap }$\epsilon
_{q}\approx \Delta $ \textit{for excitation.}

The lower energy excitation comes from the indirect path and is given by the
second pole $g\left( q,\omega \right) $ and occurs where the denominator of
the expression \ref{comact G(q,w)} goes to zero. This will happen whenever 
\begin{equation}
g_{o}\left( q,\omega \right) =\frac{1}{\Sigma \left( q,\omega \right) }
\label{condition of divergence}
\end{equation}%
If this occurs at $\omega =o,$ then the real part of $g\left( q,\omega
\right) $ has diverged leading to \ phase transition while the imaginary
part is related to the real part through expression of \ref{kramer'skronog}%
.The spectral weight of these low energy long lasting or slow response
excitations is to be identified as the bulk of our conscious experience. The 
\textit{Spectral weight \ }$\mathcal{A}\left( q,\omega \right) $ \textit{of
these excitations have a life time and come essentially from the self energy
part of the propagator }and is given by\cite{mohan}

\begin{equation*}
A\left( q,\omega \right) \approx \func{Im}\Sigma \left( q,\omega \right)
\end{equation*}

The spectral weight has the simple expression 
\begin{equation*}
A\left( \omega \right) =\sum_{\omega _{n}}\left[ \left\langle n\left\vert
\Psi ^{\dagger }\right\vert 0\right\rangle \right] ^{2}\delta \left( \omega
-\omega _{n}\right)
\end{equation*}%
This shows that external world causes real neural excitations, the delta
function in the energy summation assures energy conservation, while the
square of the matrix element gives us the intensity of the excitation
spectrum. These single excited particle states constitute bulk of the
amplitude mode. These are dissipative modes and hence lead to genuine
conscious perception process.

Our ground state defined by the minimum of Gibb's free energy at $T=T_{R},$
is where world $\Omega $ is absent (at this minimum ).This is analogous to
screening out of magnetic field by a superconductor. Our $\mathcal{I}$ sits
in this energy minimum and fluctuates out of this minimum when interacted on
by the world. This $\mathcal{I}$ has an amplitde and a phase, the unique
phase of the broken symmetry. The single particle propagation operator $%
\varphi _{c}(i,t)$ that we described describes the amplitude oscillation and
is an amplitude mode. It exists even if $\mathcal{I}$ is zero as long as $%
\left\vert \mathcal{I}\right\vert $ is non-zero. It is a high energy mode,
is an amplitude fluctuation, where creation of quasi-particle like
excitation i.e: infon particles require finite energy. These have a gap $%
\Delta $ or `mass'. We can give a number to the gap if we recall that the
critical voltage necessary to initiate action potential along an axon is
typically $\sim 100$ mv. This can be taken as the value of $\Delta .$
Because of $\theta -$symmetry breaking, there is a second fluctuation mode
in the potential well of the Mexican hat\cite{negele}. This one ( present
only when $\mathcal{I}$ $\neq 0$) is the low energy mode, due to phase $%
\theta $-fluctuation, where the order parameter fluctuates locally along the
\textquotedblleft ring of the mexican hat along the minimum
energy\textquotedblright . This phase fluctuation mode is also known as
Goldstone mode and is a phonon, whose energy is given by $\left( \text{if
the infons are charge neutral}\right) $ 
\begin{equation}
\hbar \omega _{q}=\upsilon q  \label{phonon mode}
\end{equation}%
It is a gapless collective mode where $\upsilon $ is the velocity of mode
propagation. This really just a density fluctuation and is sound wave like.
Because of the gap in the single particle excitation spectrum, the sound
wave like mode has virtually no dissipation or damping in the low energy
sector. We like to associate this mode with consciousness of thought like
processes. Here the excitation may be carried as a soliton or a sound
packet, going over a large distance adiabatically, losing no energy in the
transport. Because it is a very stable mode, it has virtually no decay
channel or imaginary part except at higher $q-vectors$ where it will merge
into the continuum of the amplitude mode and will dampen and become part of
emotional perception.We cannot be too conscious of short $q$ ( long wave
length),low $\omega $ (very low energy) thought waves ; they will remain
subliminal.On the other hand, the quasiparticle like amplitude oscillation
can have a fairly large imaginary part corresponding to real excitation but
with a life time This we believe is responsible for consciousness of
emotional proceses.

Phase and amplitude mode will couple if phase fluctuation, which is density
fluctuation couples with amplitude fluctuation, which is a single particle
excitation. There is a neural cut off at low and high energy.A violent shock
that makes $\mathcal{I}$ go over the high energy threshold is not perceived
by the mind, because precisely those regions have no spectral weight. As a
result of the shock, we may become unconscious \textit{catapulting} $I$%
\textit{\ to a metastable equilibrium a different extremum of free energy.
All conscious perception of the incident including pain,that accompanied the
intense shock,vanishes}.

In this section we have seen that cognitive response due to parrticle
excitation has two essential channels. One is the swift response, often
needed for biological survival, which is a high energy virtual excitation
process and is largely subconscious. It is instantaneous reaction and we are
barely aware of what is going on. The second channel is the slow response,
the propagator takes routes and detours, uses low energy circuits and loops,
is mainly dissipative because it is the imaginary part of overall cognitive
susceptibility.This at the root of conscious perception.

\section{Discussion}

The all - important self operator, has carved out of mind-space, cognitive
order or the $\mathcal{I-}field\mathcal{.}$ It pervades uniformly whole
space. It has given rise to a spatio-temporally homogeneous order parameter $%
\mathcal{I}$ that constutes our mental base. This $\mathcal{I}$ is the
executor of what we call, \textbf{our mind}. One cannot give a specific
neuron label to it; in order to get it, we have integrated over all the
neuron coordinates. We have asserted that This $\mathcal{I}$ and synaptic
self are identical. This is a highly questionable assertion. So far there
seems to be no concrete evidence of $\mathcal{I}$ surviving loss of personal
memory or other pathological neural disorder which seems to justify it.
Penfield seems to think the contrary. It is of interest to quote from
Penfield:\cite{penfield2} "\textit{It is what we have learnt to call the
mind that seems to focus attention. The mind is aware of what is going on.
The mind reasons and makes new decisions.It understands. It acts as though
endowed with an energy of its own. It can make decisions and put them into
effect by calling upon various brain mechanisms. It does this by activating
neuron-mechanisms.}" And he says a little further that \textit{" there is no
place in the cerebral cortex where electrical stimulation will cause a
patient to believe or to decide." }Hence one should be very cautious about
our assertion.

As we have seen $\mathcal{I}$ is also synchronous with our memory, which in
reality is a huge ($10^{11}\times 10^{11})$ matrix constituted with local
cognitive order on each and every neuron and space-time correlation between
them. When parts or whole of memory is gone, we lose our sense of the
precious $\mathcal{I}$.The global cognitive order has phase coherence
because it has got a fixed phase $\theta ,$ a different one for every brain
and which confers on each one of us,the unique personality that we have.
When we are in our ground state, at the minimum of the free energy parabola,
there is neither world or world awareness. Any fluctuation of $\mathcal{I}$
can only be local in space and time and gives rise to vastly different
excited states $\left\{ m\right\} $ of the mental space.

Operators $s_{m}$ are non-hermitian and the world they create are real but
not measurable in the physical sense. The essence of sensory experience,
named `qualia' by philosophers, that includes colour, harmony, odor and alas
pain are only too real, none measurable (not Hermitian), nor \textit{%
explainable} by the physical nature of the stimuli. When one comes to think
of it, physical properties that we attribute to things is not an intrinsic
characteristic of the outside world. These are created by $s_{m}$ operators
in the mental space.

The vacuum state $\left\vert 0\right\rangle $ on which the exponentiated
creation self operator $\psi ^{\dagger }$ acts \textit{is the pure genetic
material in the chromosomic soma of every neuron.} The operation is the
attempt by self to express and make explicit the unique physical identity of
each individual $\mathcal{I}$. This is the unconscious cognitive state 
\textit{affirming pure bodily self}, a process that must start in the womb
in the very first weeks after conception. The operator operates in
anticipation of future, prepares the representation of the body and bodily
related cognitive function in the brain. The motor area \ will be active to
help in this representation; the Penfield Homunculus\cite{markbear} map
would begin to be etched out. Sensations will follow upon birth and find
templates ready, unto which thoughts can latch into. All this is still in
the future, all this is a premonition of that future. One can almost say
that cause of all this activity is in the future, that $\mathcal{I}$ causes
itself ! It would continue long into the second year of the baby after
birth, to incorporate the varied input from the outside sensory world so as
to add the conscious\textit{\ narrative self} to the zeroth order \ bodily
self and thus complete the individuation process.

Organisms have to be understood as a mesh of virtual selves. As Varela put
it "\textit{\ I don't have one identity, I have a bricolage of identities.I
have a cellular identity, I have an immune identity, I have a cognitive
identity .}"\cite{varela}The $s_{m}$\ and its Hermitian conjugate $%
s_{m}^{\dagger }$\ operators are operators of self and as such they are
embedded into our genetic identity\textit{.} They are simply there and go on
creating a variety of instruction protocols that are needed for the brain to
be the wonderful smooth machine it is.They start acting as soon as the first
group of neurons are functional in the womb and create out of the genetic
endowment of each individual a world of representations that are previsual,
prelexique, a primordial world of ideas and sensations and categories only,
before being named or verbalised. The cognitive ground state of the baby
brain, as soon as the cognitive order parameter $\left\langle \psi
\right\rangle $ or $\mathcal{I}$ is non-zero (when it is about 2 yrs old) is
ready to interpret the outside world and to extract a coherent meaning out
of the divers exterior stimuli. From the outside world, both consciousness
and memory will form. But in the construction of $\ \mathcal{I},only$ the
genetic material is transcripted and that will serve as a template for the
world outside. Through this $\mathcal{I},$ the world within will meet the
world without.

Blocking of $\theta $ at an arbritrary value is called symmetry breaking.
This often occurs in certain class of phase transitions , where a lower
symmetry ordered phase emerges from a higher symmetry chaotic phase. In our
case emergence of $\mathcal{I}$ signifies a rupture of the multidimensional $%
U(N)$ symmetry, from objectivity to subjectivity establishing a genetic
affirmation of personality. Each $\theta $ is a different individual, a
completely different view of space-time. Blocking of global $\theta $ at
some value and that remains blocked signifies an extraordinary phase
stiffness. In order for this to happen, the infon population $N_{o}$ must be
large and vary a great deal. This number varies because brain is plastic and
the fluidity of the information flow is matched by continuous birth and
death of synaptic connections. Because the brain is an open system, open to
the world, the information content as well as their number is a continually
fluctuating quantity. This flux and influx of information is precisely the
condition necessary to achieve a phase coherent state. The information must
fluctuate a great deal around some average value which permits brain to
achieve phase coherence between different parts and we can extract a
coherent meaning from our sensory input. Nothing prevents $\theta $ to
fluctuate locally and give rise to excitations in the mind which are mind
waves. These excitations could be collective and massive extended through
the whole system as in an epileptic seizure or could be single particle
like, intense and localised, like spikes of pain. Importance is maintaining
the phase coherence, no matter what and in this $\mathcal{I}$ is both
witness and regulator of coherence and assures a maximum of information
flow, including contradictory information so as to create the overall
meaning. The traffic exchange between different neurons through the synaptic
clefts \ is a key player in this game. Nothing is more eloquent in this
respect than the behaviour of the two hemispheres of the brain, left and
right . The left brain is analytical, logical, time sensitive, while the
right processes information in a holistic way rather than breaking them down
and more involved with sensory perception rather than abstract cognition.
Between the two hemispheres is a thick bundle of axons or nerve fibers,
about 80 million called \textit{corpus callosum} \ that handles the heavy
traffic of information without which we shall not get a global conscious
coherent state. If this traffic is interrupted, personality disorder will
arise, and most likely two different coherent states, one on the left and
another one on the right will rise and exist side by side.\ Symmetry
breaking into more than one $\Theta $ is conceivable in certain cases of
brain disorder where the free energy of the two $\Theta $-states being the
same, the $\psi -$operator will flip-flop between two equivalent metastable
equilibrium and the resultant personality will effortlessly slip from one
into other \textit{but with the same sense of }\textquotedblleft $\mathcal{I}
$\textquotedblright $.$ Here we may quote the noted neurologist Ramachandran 
\cite{ramachandran} who writes " The sense of `unity' of self also desrves
comment. Why do you feel like `one' despite being immersed in a constant
flux of sensory impressions, thoughts and emotions? .....Perhaps the self by
its very nature can be experienced only as a unity." And a little further "
Even people with so-called multiple personality disorder don't experience
two personalities simultaneously--- the personalities tend to rotate and are
mutually amnesic".

The brain order parameter $\Phi _{eq}$, at the free energy minimum
represents the lowest energy state of the cognitive system. The order
parameter $\Phi $ represents a whole landscape of free energy valleys and
hills (different states of awareness) rather than one absolute minimum. 
\textit{The }$\mathcal{I}$ \textit{that emerges is a tremendous transition
from the\ Self that is simply an operator }$\psi $\textit{\ to what becomes}%
\ $\mathcal{I}$ \textbf{am}. This $\mathcal{I}$ can be thought as a self
appointed instructor of the cognitive machine: the $\mathcal{I}$ that lives,
governs and presides over our thought, action, emotions and our dreams.

We want to make a comment here about Dream state. If from a state of
consciousness, the organism enters rapidly into sleep, world would not have
had time to be totally expelled or annealed out, before falling asleep. This
remanence of the world, these trapped flux of world-lines, resemble trapped
magnetic flux in a superconductor as it is cooled in a magentic field, and
may be the cause of vivid dreams. These dream states cannot be eliminated
and the system will oscillate between deep dreamless ground state of sleep
and patches of dream where local neuronal excitations continue to persist.

Before ending this discussion, a word may be apropriate about these self
operators we have employed. Sakurai \cite{sakurai} had written \`{a} propos
the creation, destruction and preservation operators used in quantum
mechanics that these \ \textit{"three operators correspond \ respectively to
the Creator (Brahma), the Destroyer (Siva), and the Preserver (Vishnu) in
Hindu mythology."} If anything the operator of cognitive Self $s_{m}$ fits
perfectly this description. Self creates, self destroys, self also
preserves. Between this triad of operators, $S=\left\{ s_{m}^{\dagger
},s_{m},n\right\} $ that we may designate by the symbol $S$, the whole human
drama is enacted.

\bigskip

\section{Appendix}

We shall give here a simple model hamiltonian that captures the role of
synaptic connectivity to bring about global consciousness response when a
single neuron gets connected to other neurons. We borrow for the purpose the
simple tight binding hamiltonian of electrons from solid state physics\ref%
{ashcroft} .

The response of a single neuron $i$ , \ called \ $\chi _{i}^{o}$ is defined
as (superscript zero, signifying zeroth order)response function in the
absence of external perturbation 
\begin{equation}
\chi _{i}^{o}\left( t\right) =\left\langle \psi _{i}\left( t\right) \psi
_{i}^{\dag }\left( o\right) \right\rangle \text{ ; }\chi _{i}^{o}\left(
\omega \right) =\int_{-\infty }^{\infty }dt\text{ }\chi _{i}^{o}\left(
t\right) \exp i\omega t  \label{singleneuronsusceptibility}
\end{equation}%
As we have already expressed, in the presence of \textit{external force} $%
F_{i},$ acting on the neuron $i$ we can write 
\begin{equation*}
\left\langle \psi _{i}\left( \omega =0\right) \right\rangle =\chi
_{i}^{l}\left( \omega =0\right) F_{i}
\end{equation*}

Here $\chi _{i}^{l}\left( \omega =0\right) $is the full interacting local
susceptibility of the single neuron, when it is giving and receiving signals
to and from all other neurons. First we write down the simplest hamiltonian
we can \ that catches the essential dynamics of information transfer between
neurons and also between neurons and the world.

This is written as sum of three essential parts

\begin{eqnarray}
H_{n} &=&\sum_{i}\epsilon _{i}n_{i}-\mu
\sum_{i}n_{i}+\sum_{i,j}V_{ij}n_{i}n_{j}  \label{hamiltonian} \\
H_{t} &=&-\sum_{i,j}\left( T_{ij}\psi _{i}^{\dagger }\psi _{j}+h.c\right) \\
H_{ext} &=&\sum_{i}g_{i}\left( \Omega _{i}\psi _{i}^{\dagger }+\Omega
_{i}^{\ast }\psi _{i}\right)
\end{eqnarray}

\textit{Here }$H_{n}$ is the hamiltonian that has onsite site energy $%
\epsilon _{i},$chemical potential $\mu $ of infon on each site $i$ as well
as some assumed repulsive energy between neuron population at sites $i$ and $%
j$ The part of the hamiltonoian $H_{0}+H_{t},$ when written for bosons is
well-known. In the special case, when $V_{ij}$ is repulsive and if $%
V_{ii}=\infty ,$no two bosons can occupy the same site (hard core limit).
The lattice hamiltonian we used, in the hard core boson limit in
translationally invariant lattice\ is well -known to posess a superfluid
ground state. \cite{matsubara} Neuron network in human brain is highly
irregular, is plastic, the synaptic interconnections are far from being
identical and continually evolving. Any conclusion about its superfluidity
should await a long time until we can have clean non-invasive experimental
data.

The all important tunneling of information from neuron $i$ to neuron $j$ is
given by the tunneling (also called hopping) matrix element $T_{ij}$ through
the synapses in between. The expression $h.c$ within the bracket signifies
the reverse or hermitian conjugate process of info-transfer from $j$ to $i.$
The term $H_{ext}$ of the hamiltonian expresses interaction of the neuron
with the external world. This includes one's own body exterior to the
cognitive system as well as the world around. The first three terms can be
written in the Hartree form \ as pure onsite part . We thus divide the
Hamiltonian in two parts, 
\begin{subequations}
\begin{eqnarray}
H_{n} &=&\sum_{i}H_{i}+H_{int}  \label{tunnelinghamiltonian} \\
H_{i} &=&\epsilon _{i}n_{i}-\mu n_{i}+V_{H}n_{i} \\
where\text{ \ }V_{H}\text{\ } &=&\sum_{j}V_{ij}\left\langle
n_{j}\right\rangle \\
H_{int} &=&-\sum_{i,j}(T_{ij}\psi _{i}^{\dagger }\psi
_{j}+h.c)+\sum_{i}g_{i}\left( \Omega _{i}\psi _{i}^{\dagger }+\Omega
_{i}^{\ast }\psi _{i}\right)  \notag
\end{eqnarray}

The term$V_{H}$ is the Hartree term and has been absorbed into the site
energy $i$ . The term $H_{int}$ contains interaction with other neurons and
with the world. The all-important tunneling Hamiltonian will be simplified
as 
\end{subequations}
\begin{equation}
H_{t}=-\left[ \sum_{i,j}T_{ij}\left\langle \psi _{j}\right\rangle \psi
_{i}^{\dag }+\sum_{i,j}T_{ij}\left\langle \psi _{i}^{\dag }\right\rangle
\psi _{j}\right] +\sum_{i,j}T_{ij}\left\langle \psi _{i}^{\dag
}\right\rangle \left\langle \psi _{j}\right\rangle +h.c
\label{averagetunneling}
\end{equation}

The first two terms of the equation \ref{averagetunneling} act like a
molecular field on the information operators at $i$ $\ \&$ $j.$The last term
is just a c-number that we neglect since it does not have any operator
character .We want to express the interaction hamiltonian into a molecular `%
\textit{Weiss Field}' acting on the site $i$. We first consider just nearest
neighbor tunneling to get an order of magnitude idea of the effect of the
molecular field of nearest neighbors \ or short range tunneling on the \
static ($\omega =o$) single neuron susceptibility \ This is given by

\bigskip 
\begin{equation}
H_{t}^{o}=-\nu T_{nn}\left[ \left\langle \psi _{j}\right\rangle \sum_{i}\psi
_{i}^{\dag }+\left\langle \psi _{j}^{\dag }\right\rangle \sum_{i}\psi _{i}%
\right] +h.c
\end{equation}

Here $\nu $ is the number of first near-neighbor neurons $\left( \sim
10^{4}\right) $ of a given neuron connected through synapses, with an
average tunneling amplitude $T_{nn}.$ $T_{nn}$has the dimension of
energy.Thus the tunneling term gives a Weiss molecular field contribution
acting on the site $i$%
\begin{equation*}
F_{t}=-\nu T_{nn}\left\langle \psi _{j}\right\rangle
\end{equation*}%
Similarly external world acts with a \textit{`force'}%
\begin{equation*}
F_{ext}=-g\Omega _{i}
\end{equation*}%
This permits us to write%
\begin{equation*}
\left\langle \psi _{i}\right\rangle =\chi _{i}^{o}\left[ g\Omega _{i}+\nu
T_{nn}\left\langle \psi _{j}\right\rangle \right]
\end{equation*}%
We make now the homogenity assumption $\left\langle \psi _{j}\right\rangle
=\left\langle \psi _{i}\right\rangle $ and write a mean-field susceptibility 
\begin{equation*}
\left\langle \psi _{i}\right\rangle =\chi _{i}^{l}\text{ }\Omega _{i}
\end{equation*}%
The R.P.A or mean-field \ interacting susceptibility is now expressed in the
compact form%
\begin{equation}
\chi _{i}^{l}=\frac{\chi _{i}^{o}}{1-\nu T_{nn}\chi _{i}^{o}}
\label{local cognitive response}
\end{equation}%
For a `free' particle like behaviour of infons in the symmetry unbroken
phase, we may write the \ real and imaginary part (as a Hilbert transform of
the real part) of the non-interacting susceptibility as 
\begin{eqnarray*}
\func{Re}al\text{ }\chi _{i}^{o} &\approx &\rho _{o} \\
\func{Im}\chi _{i}^{o} &\approx &\rho _{o}\omega \tau
\end{eqnarray*}%
Here $\rho _{o}$ is density of states of the infons (number of infons per
unit energyper unit volume) as $\omega \rightarrow 0,$ and $\tau $ is a
characteristic relaxation time of the excitations, assumed frequency
independent. Now we can equate the real and imaginary part of the
interacting susceptibility of expression \ \ref{local cognitive response}
and obtain 
\begin{equation}
\func{Re}al\text{ }\chi _{i}^{l}=\frac{\rho _{o}-\lambda \rho
_{o}^{2}(1+\omega ^{2}\tau ^{2})}{\left( 1-\lambda \rho _{o}\right)
^{2}+\lambda ^{2}\rho _{o}^{2}\omega ^{2}\tau ^{2}}
\label{Real part cognitive susc}
\end{equation}%
The imaginary part is given by 
\begin{equation}
\func{Im}\chi _{i}^{l}=\frac{\rho _{o}\omega \tau }{\left( 1-\lambda \rho
_{o}\right) ^{2}+\lambda ^{2}\rho _{o}^{2}\omega ^{2}\tau ^{2}}
\label{Imaginary part cognitive susc}
\end{equation}%
Here we have wriiten the symbol $\lambda $ for a characteristic energy
parameter of synaptic connectivity, 
\begin{equation*}
\lambda =\nu T_{nn}
\end{equation*}%
The real part of interacting susceptibility as $\omega \rightarrow 0$ blows
up as $\lambda \rho _{o}\rightarrow 1.$This gives us the \ critical value of
neuronal connectivity when \ $\rho _{o}=\frac{1}{\nu T_{nn}}.$\textit{This
infinity \ signifies an unstability and a phase change indicating a new
cognitive state for the child, that of self consciouness developing rapidly
out of consciousness}. The phenomenon has a great degree of similitude to
superconductive instability\cite{benoy}. This is precisely the point where $%
A,$ the coefficient of the second order term in the Ginzburg- Landau
expression of the preceding section, goes to zero $\left( 2A=\frac{1}{%
\sum_{i}\chi _{i}^{l}\left( \omega =0\right) }\right) $ . From this point
onwards, $A$ \ can be negative, free energy function races to a stable
minimum at the non-zero value of $\Phi _{eq}$. $\mathcal{I}$ \ \textit{can
emerge as a self conscious self.}

\bigskip

\section{Conclusion}

In conclusion we can summarise our investigation of Consciousness as a three
step approach :

First and foremost we have defined Mind as a quantum field whose excitations
are called quanta of information .

Second, we have defined a quantum operator $S$ representing self, whose
action on the mind vacuum state called $\left\vert 0\right\rangle $
generated a coherent macroscopic functional space of mind where a non-zero
average of the self operator emerged as $\mathcal{I}$. This $\left\vert 
\mathcal{I}\right\rangle $ field replaces the original vacuum $\left\vert
0\right\rangle $ state and is our personal ground state of the mind.

Finally, energy excitations out of this ground state, as a result of
interaction with outside world, is perceived by $\mathcal{I}$ as being
conscious of the world. Consciousness is defined as a causal response
function that vanishes when one is in the true ground state $\left\vert 
\mathcal{I}\right\rangle .$

\bigskip

\textbf{Acknowledgement }

\textit{I want to acknowledge my deep indebtedness to Philippe Nozi\`{e}res
\ of College de France for teaching \ some of us over the years many aspects
of quantum fluid including Bose-Einstein condensation and superfluidity. I
am also grateful to professor Jean \ Perret, former head of the department
of Neurology at University Joseph Fourier, Grenoble for his series of
lectures on human brain.}

Many persons contributed through their discussions with me. I want to thank
particularly Prof Ferdinando de Pasquale of University of Rome, La Sapienza,
dept of Physics, for bringing to my attention analogies with Quantum
decision theory, to Dr Amitabha Chakrabarti of Ecole Polytechnique, Paris
for very perceptive comments about emergence of `I', to Dr Mario Cuoco of
Physics Dept, University of Salerno, to Prof T.V.\ Ramakrishnan of The
Physics Dept, University of Benaras, India for pointing out importance of
Self in Indian Philosophy. Thanks are also due to Prof Philippe Nozieres for
questioning quantum nature of brain processes and to Dr Timothy Ziman of
Theoretical Physics, Institut Laue-Langevin, Grenoble for suggesting that
emergence of consciousness may have to do with time scales. Finally Jeanine
and Kolyan Chakraverty bore the brunt of many interrogations on the subject
with me these past years, shared some of their insights and I am grateful
for that.


\begin{thebibliography}{99}
\bibitem{hopfield} J.J.Hopfield , Rev Mod Phys, 71,431-437 (1999)

\bibitem{amit} D.Amit " Modeling Brain Functions " (Cambridge University
Press, Cambridge, 1989).

\bibitem{haken} H.Haken " Brain Dynamics " (Springer, Berlin 2008)

\bibitem{bohr} N.Bohr, \ Naturwissenschaft, 17, 483-4486; also " Light and
Life", Nature 131,421-423 (1933).

\bibitem{rpenrose} R.Penrose, " The Emperor's New Mind " (Oxford University
Press, Oxford 1989).

\bibitem{satinover} J.Satinover " The Quantum Brain " (Wiley, New York,
2001).

\bibitem{frohlich} H.Frohlich \ Phys Lettr, A, 402 , 26, (1968)

\bibitem{beck} F.Beck \& J.Eccles, Proc NAt.\ Acad. Sci, U.S.A , 89,11357
(1992).

\bibitem{vitiello} Giuseppe Vitiello " My Double Unveiled " (John Benjamins
Publ, Amsterdam, 2001), Chapter 5.

\bibitem{yukalov} I.Yukalov and Didier Sornette, ar Xiv /0802.3597v3 $\left[ 
\text{physics.soc-ph}\right] $ 17 march, 2010 and references therein.

\bibitem{penfield} W.Penfield \& P.Perot, Brain 86 (1963) 595-696

\bibitem{timgreen} T.Green, Stephen Heinemann \& Jim Gusella, Neuron, (1998)
, 420

\bibitem{markbear} M.F.Bear, B.W.Connors \& M.A.Paradiso, "NEUROSCIENCES"
(Lippincourt Williams \& Williams),23

\bibitem{descartes} A. R. Damasio "Descartes' Error" (Quill, Harper Collins
Publishers, New York,2000) p123.

\bibitem{sherrington} C.Sherrington "Man On His Nature " (Cambridge
University Press 1963)

\bibitem{vonneumann} Von J. Neumann, " Mathematical Foundations of Quantum
Mechanics" (Princeton University Press, Princeton, 1955).

\bibitem{monkey} M.A.L. Nicolelis et al, Jl of Neuroscience, 25 (2005),4681;
The New York \ Times, page 6, section " Science \& Technology ", 26th
january, 2008

\bibitem{glauber} R.J.Glauber, Phys Rev 130 (1963),25296 (2005)

\bibitem{cabannis} P.Cabannis, " Trinit\'{e} de Physique et Moral de L'Homme
\textquotedblleft\ Second Memoir \textquotedblright\ (1802)

\bibitem{wigner} T.D.Newton and E.P.Wigner, Rev Mod Phys, 21 (1949) 400

\bibitem{feynman} R.P.Feynman and A.R.Hibbs, " Path Integrals and Quantum
Mechanics " McGraw Hill, New York, 1968.

\bibitem{Dirac} P.A.M Dirac, The Principles of Quantum Mechanics (Clarendon
Press, Oxford, 1958).

\bibitem{Bose} S.N.Bose, Z.Phys,26,(1924) 178

\bibitem{Negel} Thomas Negel , " The View from Nowhere" (Oxford University
Press, 1986).

\bibitem{walecka} Alexander Fetter \& John Walecka, " Quantum Theory of Many
Particle Physics " (Mcgraw -Hill Book Company ,Newyork,1971) 12.

\bibitem{wolf} Leonard Mandel \& Emil Wolf " (Optical Coherence \& Quantum
Optics " , (Cambridge University Press, 2008),522

\bibitem{ledoux} Joseph Ledoux " Synaptic Self "( Penguin Books,2002),67.

\bibitem{coleman} S.Coleman " Aspects Of Symmetry " (Cambridge University
Press, Cambridge, 1985), chapter 5.

\bibitem{shear} Jonathan Shear (ed) " Explaining Consciousness ---The 'Hard
Problem' " (MIT Press,1997).

\bibitem{chalmers} David J. Chalmers, " The Conscious Mind " (Oxford
University Press,)1996.

\bibitem{nozieres} David .Pines \& Philippe Nozi\`{e}res, \ " Theory of
Quantum liquids", Vol 1, (Addison-Wesley , Newyork, 1988), 95.

\bibitem{degennes} P.G.de Gennes " Superconductivity of Metals and Alloys
"(Addison-Wesley, Newyork 1989),171.

\bibitem{negele} John Negele \& Henri Orland " Quantum many Particle Systems
" (Addison Wesley, Newyork 1988), 217.

\bibitem{Bloch} C.Bloch \& C.deDominicis - Nucl Physics 7 (1958) 459.

\bibitem{mohan} Gerald D. Mahan, " Many-Particle Physics " \ (Plenum Press,
Newyork, 1990) 145.

\bibitem{penfield2} Wilder Penfield, " Mystery Of The Mind " (Princeton
paperbacks, Princeton University press,1978) 75.

\bibitem{varela} Francisco Varela, " The Third Culture: Beyond the
Scientific Revolution " by John Brockmann, chapter 12 (Simon \& Scuster,
1995).

\bibitem{ramachandran} V.S.Ramachandran " The Emerging Mind " - The Reith
Lectures, 2003 (Profile Books, London) 123.

\bibitem{penrose} Roger Penrose, " Shadows of The Mind " ( Oxford University
Press, 1994 ) 204

\bibitem{philipps} Philip Philipps \& Denis Dalidovitch, SCIENCE vol 302,
(2003),243.

\bibitem{sakurai} J.J.Sakurai, " Advanced Quantum Mechanics" \ (Benjamin/
Cummings Publishing, California) 27.

\bibitem{ashcroft} Neil W. Ashcroft \& N.David Mermin-- " Solid State
Physics "

\bibitem{matsubara} T.Matsubara \& H.Matsuda, Progr Theor Physics 16 (1956)
569.

\bibitem{benoy} B.K.Chakraverty, Phys Rev B, 48 (1993) 4047.
\end{thebibliography}
\end{document}